\documentclass[12pt]{article}

\usepackage[dvipdfmx]{graphicx}
\usepackage{amssymb}
\usepackage{amsmath}
\usepackage[all]{xy}

\def\IN{\mathbb {N}}
\def\IZ{\mathbb {Z}}
\def\IQ{\mathbb {Q}}
\def\IR{\mathbb {R}}
\def\IC{\mathbb {C}}

\def\ICP{\mathbb {CP}}
\def\IRP{\mathbb {RP}}

\renewcommand{\thefootnote}{\fnsymbol{footnote}}

\makeatletter
 \renewcommand{\theequation}{%
       \thesection.\arabic{equation}}
 \@addtoreset{equation}{section}
\makeatother

\makeatletter
\def\eqnarray{%
 \stepcounter{equation}%
 \let\@currentlabel=\theequation
 \global\@eqnswtrue
 \global\@eqcnt\z@
 \tabskip\@centering
 \let\\=\@eqncr
 $$\halign to \displaywidth\bgroup\@eqnsel\hskip\@centering
 $\displaystyle\tabskip\z@{##}$&\global\@eqcnt\@ne
 \hfil$\displaystyle{{}##{}}$\hfil
 &\global\@eqcnt\tw@$\displaystyle\tabskip\z@{##}$\hfil
 \tabskip\@centering&\llap{##}\tabskip\z@\cr}
\makeatother

\begin{document}
\begin{titlepage}

\begin{center}
\vspace*{1cm}
{\Large \bf
Open Mirror Symmetry for Higher Dimensional\\[0.8em]
Calabi-Yau Hypersurfaces}
\vskip 1.5cm
{\large Yoshinori Honma${}^{a}$\footnote[2]{honma@itp.uni-leipzig.de} and Masahide Manabe${}^b$\footnote[3]{masahidemanabe@gmail.com}}
\vskip 1.0em
{\it 
${}^a$%
Institut f$\it \ddot{\textrm{u}}$r Theoretische Physik, Universit$\it \ddot{\textrm{a}}$t Leipzig \\
Br$\it \ddot{\textrm{u}}$derstrasse 16, Leipzig, D-04103, Germany\\

${}^b$%
Faculty of Physics, University of Warsaw\\
ul. Pasteura 5, 02-093 Warsaw, Poland\\}
\end{center}
\vskip2.5cm

\begin{abstract}
Compactifications with fluxes and branes motivate us to study various enumerative invariants of Calabi-Yau manifolds. In this paper, we study non-perturbative corrections depending on both open and closed string moduli for a class of compact Calabi-Yau manifolds in general dimensions. Our analysis is based on the methods using relative cohomology and generalized hypergeometric system. For the simplest example of compact Calabi-Yau fivefold, we explicitly derive the associated Picard-Fuchs differential equations and compute the quantum corrections in terms of the open and closed flat coordinates. Implications for a kind of open-closed duality are also discussed.
\end{abstract}
\end{titlepage}


\renewcommand{\thefootnote}{\arabic{footnote}} \setcounter{footnote}{0}

\section{Introduction}\label{sec:intro}

Mirror symmetry is an efficient tool to compute the enumerative invariants of Calabi-Yau manifolds such as Gromov-Witten invariants as first demonstrated in \cite{Candelas:1990rm} to the quintic threefold. The genus zero Gromov-Witten invariants are equivalent to the number of holomorphic spheres in the manifold, and it is quite natural to ask whether one can consider more general enumerating problem in the framework of mirror symmetry. One of these attempts has been undertaken in \cite{Walcher:2006rs} for the case of holomorphic disks in the compact Calabi-Yau manifold, and the disk partition function for the quintic threefold was clarified (see also \cite{Solomon:2006i}). This open mirror symmetry prescription was subsequently developed in \cite{Krefl:2008sj, Knapp:2008uw, Walcher:2009uj} for other one- and two-parameter Calabi-Yau threefolds.
By extending the holomorphic anomaly equation for the closed topological strings \cite{Bershadsky:1993cx}, the higher genus invariants for open topological strings have been also studied \cite{Walcher:2007tp, Cook:2007dj, Bonelli:2007gv, Neitzke:2007yw}. More mathematical treatment of disk enumeration can be found in \cite{Pandharipande:2006d, Fukaya:2009c} for the A-model and \cite{Morrison:2007bm} for the mirror B-model.

As described in \cite{Walcher:2006rs}, in order to derive the disk partition function of the open topological string, it is sufficient to introduce the branes from the involution on the Calabi-Yau manifold.
However, this involution brane does not have an explicit open string moduli dependence and therefore the structure of open string deformation cannot be extracted from this ``on-shell" formalism. Based on the earlier works of \cite{Lerche:2002ck, Lerche:2002yw}, the incorporation of open string moduli was carried out in \cite{Jockers:2008pe} for the case of compact Calabi-Yau threefolds.
By studying the associated Hodge structure, they derived the Picard-Fuchs differential equations depending on both open and closed string moduli parameters.

Thereafter, using the toric geometry prescription, an alternative and more efficient method to derive the equivalent differential system was constructed in \cite{Alim:2009rf}.
Interestingly, this formulation implies that there exists a duality between the off-shell open topological strings on a Calabi-Yau threefold with branes and the closed topological string on a Calabi-Yau fourfold without branes \cite{Mayr:2001xk, Lerche:2001cw, Alim:2009bx, Aganagic:2009jq, Jockers:2009ti}. 

Therefore the natural question is whether one can construct the off-shell formalism and recognize signs of a similar duality in higher dimensions.
On-shell mirror symmetry prescription and the direct calculation using the localization formula have been argued in a certain class of higher dimensional Calabi-Yau manifolds in several contexts.
However, to the best of our knowledge, an explicit generalization including the open string deformation has not been 
elucidated in dimensions greater than three. The aim of this paper is to provide a first step to clarify this fascinating issue, which may have a wide range of applications. For instance, in the case of Calabi-Yau fivefolds, our prescription might be useful to compute the non-perturbative corrections in the effective superpotential of the M-theory compactification studied in \cite{Haupt:2008nu}.

This paper is organized as follows. First we take a look at the disk enumeration problem from the aspect of mirror symmetry in Section \ref{sec:disk}. It gives a brief introduction to the closed and on-shell open mirror symmetry in general dimensions. In Section \ref{sec:disk_GD}, we describe how to incorporate the open string deformation into the present framework and construct the necessary ingredients via Hodge theoretical approach. Furthermore, we explicitly derive a Picard-Fuchs system associated with the open and closed string deformations for an example of compact Calabi-Yau fivefold. In Section \ref{sec:disk_glsm}, we present another efficient formulation in terms of the toric geometry and show that the resulting differential system is equivalent to the one obtained in Section \ref{sec:disk_GD}. Finally, in Section \ref{sec:h_open_closed}, we will also refer to the implications of the off-shell formalism for a certain kind of open-closed duality. Section \ref{sec:conclusion} is devoted to the conclusion and discussion. We summarize the localization formulas for the related A-model computations in Appendix \ref{app:localization}. Monodromies under the analytic continuation in the moduli space is also discussed in Appendix \ref{app:monodromy}. In Appendix \ref{app:disk_two}, we discuss about disk two-point functions, and the differential equations derived in Section \ref{sec:disk_glsm} is solved in Appendix \ref{app:sol_ext_gkz}.

\section{Disk enumeration and open mirror symmetry}\label{sec:disk}

In this section, we describe the disk enumeration problem on complex odd dimensional Calabi-Yau hypersurfaces from the viewpoint of mirror B-model.

Let us consider a Calabi-Yau $m$-fold $X_{m+2}\subset {\ICP}^{m+1}$ defined by a degree $m+2$ Fermat hypersurface $\sum_{i=1}^{m+2}x_i^{m+2}=0$ in a projective space ${\ICP}^{m+1}$. Its mirror manifold $Y_{m+2}\subset {\ICP}^{m+1}$ is described by a resolution of a $({\IZ}_{m+2})^m$ orbifold of the hypersurface
\begin{equation}
Y_{m+2}:\ \ P(\psi)\equiv \sum_{i=1}^{m+2}y_i^{m+2}-(m+2)\psi\prod_{i=1}^{m+2}y_i=0,
\label{cym_hyp_def}
\end{equation}
in ${\ICP}^{m+1}$. Here $\psi$ denotes the complex structure modulus associated with the closed string deformation. The large complex structure point and the Landau-Ginzburg point in the moduli space are given by $\psi=\infty$ and $\psi=0$, respectively. The $({\IZ}_{m+2})^m$ orbifold which preserves the hypersurface constraint (\ref{cym_hyp_def}) is defined by
\begin{align}
{\IZ}_{m+2}^{(i)}:\ \ &[\ y_1:\ldots:y_{i-1}:y_i:y_{i+1}:\ldots:y_{m+1}:y_{m+2}\ ]\nonumber\\
\mapsto\ \ &[\ y_1:\ldots:y_{i-1}:\rho y_i:y_{i+1}:\ldots:y_{m+1}:\rho^{m+1}y_{m+2}\ ],\ \ i=1,\ldots,m+1,
\label{Morb}
\end{align}
where $\rho= e^{2\pi i/(m+2)}$. Note that there are only $m$ independent identifications due to the existence of overall rescaling induced by ${\IZ}_{m+2}^{(1)} \times {\IZ}_{m+2}^{(2)} \times \cdots \times {\IZ}_{m+2}^{(m+1)}$. 

Next we consider the inclusion of branes.
On the A-model side, we further introduce an involution brane on the Lagrangian submanifold $L_{m+2}=X_{m+2}^{\IR}$ in $X_{m+2}$ as the real locus which is defined by the fixed points of an anti-holomorphic involution
\begin{equation}
[\ x_1:x_2:\ldots:x_{m+2}\ ] \mapsto [\ \overline{x}_1:\overline{x}_2:\ldots:\overline{x}_{m+2}\ ].
\label{inv_brane1}
\end{equation}
on ${\ICP}^{m+1}$.
Since $L_{m+2} \cong {\IRP}^m$, we find $H_1(L_{m+2},{\IZ})\cong {\IZ}_2$. 
This means that there are two choices for specifying the involution brane. 
In the A-model on $X_{m+2}$, let us consider the worldsheet $n$-point function on the disk $D$ as
\begin{equation}
\left<{\cal O}_{h^{p_1}}\cdots{\cal O}_{h^{p_n}}\right>_D^{{X}_{m+2}},
\label{ncorr}
\end{equation}
where the boundary $\partial D$ is mapped to $L_{m+2}$. Topological observables ${\cal O}_{h^i}$ of the A-model are defined from the hyperplane class $h$ of $X_{m+2}\subset {\ICP}^{m+1}$.
This $n$-point function can be regarded as a generating function of open Gromov-Witten invariants associated with the holomorphic disks 
ending on $L_{m+2}$.\footnote{Direct calculation in the A-model via localization formula is briefly summarized in Appendix \ref{app:localization}.} Note that (\ref{ncorr}) is nonzero only if
\begin{equation}
\sum_{i=1}^np_i=\frac{m-3}{2}+n,
\label{disk_corr_cond}
\end{equation}
according to the dimensional analysis. In what follows, we concentrate on the case where
\begin{equation}
m=2\mathfrak{m}+1, \ \ \ \ \  \mathfrak{m} \in {\mathbb{N}}.
\end{equation}
Instead of (\ref{inv_brane1}), by using the automorphism group 
of ${\ICP}^{m+1}$, the Lagrangian 
submanifold $L_{m+2}$ can be also defined from fixed points under the identification
\begin{equation}
x_{2i-1}\ \mapsto\ \overline{x}_{2i},\ \ \
x_{2i}\ \mapsto\ \overline{x}_{2i-1},\ \ \
x_{2\mathfrak{m}+3}\ \mapsto\ \overline{x}_{2\mathfrak{m}+3},\ \ \ \
i=1,\ldots,\mathfrak{m}+1.
\label{inv_brane2}
\end{equation}

The efficient way to compute the open Gromov-Witten invariants is to make use of the open mirror symmetry.
In \cite{Morrison:2007bm}, by using the matrix factorization \cite{Herbst:2008jq}, a set of holomorphic two-cycles in $X_3$ wrapped by the B-brane mirror to the involution brane on $L_{3}$ was clarified. Analogously, the mirror geometry of the Lagrangian submanifold $L_{m+2} \subset Y_{m+2}$ can be described by $2\mathfrak{m}$-dimensional two isolated holomorphic curves
\begin{align}
C_{\pm}:\ \ 
&
y_{2i-1}+y_{2i}=0,\ \ \
i=1,\ldots,\mathfrak{m}+1,
\nonumber\\
&
y_{2\mathfrak{m}+3}^{\mathfrak{m}+1}\pm \sqrt{(-1)^{\mathfrak{m}-1}(m+2)\psi}\prod_{j=1}^{\mathfrak{m}+1}y_{2j-1}=0.
\label{hol_2m_cyc}
\end{align}
It would be interesting to derive this expression from the matrix factorization.

Let us consider a chain integral
\begin{equation}
\mathcal{T}_C=\int_{\Gamma}\Omega,
\label{h_dom_ten}
\end{equation}
where $\Omega$ is the holomorphic $m$-form on the mirror Calabi-Yau $Y_{m+2}$ depending on the complex structure modulus $\psi$. $\Gamma$ is an $m$-chain with the homologically trivial boundary $\partial \Gamma=C_{+}-C_{-}$, which means $[C_{+}-C_{-}]=0\in H_{2\mathfrak{m}}(Y_{m+2},{\IZ})$. 
Clearly the cycle $C_{+}-C_{-}$ defined from (\ref{hol_2m_cyc}) satisfies this condition for all $\psi$.
This quantity is a higher dimensional analog of the tension of a BPS domainwall \cite{Witten:1997ep} 
realized by a brane wrapped on $\Gamma \subset Y_3$. 
In the case of Calabi-Yau threefold, it is well understood that this chain integral is related to the D-brane superpotential 
for the four dimensional ${\mathcal{N}}=1$ supersymmetric compactifications (see, for example \cite{Baumgartl:2010ad, Grimm:2010gk} and references therein). Therefore, our result for Calabi-Yau fivefold is naturally expected to provide the ``M-brane superpotential'' in the super-mechanics theory considered in \cite{Haupt:2008nu}.

The integral (\ref{h_dom_ten}) is regarded as the on-shell disk partition function whose expression in the vicinity of the large complex structure point is given by \cite{Jinzenji:2011vm}
\begin{equation}
\tau(z)= \frac{c}{2^{\frac{m-1}{2}}}\Gamma\Big(\frac12\Big)^{m+1} \sum_{d=1}^{\infty}\frac{\Gamma\left((m+2)d-\frac{m}{2}\right)}{\Gamma\left(d+\frac{1}{2}\right)^{m+2}}z^{d-\frac{1}{2}},
\label{s_pot_gen_m}
\end{equation}
where $c$ is a normalization constant\footnote{In Appendix \ref{app:monodromy}, we see that this constant can be fixed by monodromy analysis as performed in \cite{Walcher:2006rs}.}, and
\begin{equation}
z=\frac{1}{\big((m+2)\psi\big)^{m+2}}
\label{z_psi_rel}
\end{equation}
is a local coordinate around the large complex structure point.
The disk partition function (\ref{s_pot_gen_m}) satisfies the inhomogeneous Picard-Fuchs equation associated with $C_{+}-C_{-}$ given by \cite{Walcher:2006rs, Pandharipande:2006d, Morrison:2007bm} (see also \cite{Fuji:2010uq, Walcher:2012zk, Laporte:2012hv, Jefferson:2013vfa})
\begin{equation}
{\cal L}\tau(z)=\frac{c(m+2)!!}{2^{m}}\sqrt{z},\ \ \ \ \ \ {\cal L}\equiv \theta_z^{m+1}-(m+2)z\prod_{k=1}^{m+1}
\big((m+2)\theta_z+k\big),
\label{on_inhom_PF}
\end{equation}
where $\theta_z= z \partial/\partial z$.
The solutions to the homogeneous Picard-Fuchs equation
\begin{equation}
{\cal L}\Pi(z)=0
\label{closed_PF}
\end{equation}
give the periods of the holomorphic $m$-form $\Omega$ on the mirror manifold $Y_{m+2}$,
which can be expressed by integration over the integer homology group\footnote{In general Calabi-Yau manifolds with complex dimension greater than three, we need to restrict the middle dimensional homology basis to lie in the primary horizontal subspace of homology. This restriction is a characteristic property of mirror symmetry in higher dimensions \cite{Greene:1993vm}. However, in our simple one parameter Calabi-Yau hypersurfaces $Y_{m+2}$, this restriction becomes trivial.} as
\begin{align}
{\Pi}_p(z)=\int_{\Gamma_p}\Omega(z), \ \ \ \ \
\Gamma_p \in H_m(Y_{m+2},{\IZ}),
\label{cper}
\end{align}
where $p=0,\ldots,m$. Solving the Picard-Fuchs equation (\ref{closed_PF}), we obtain the closed string periods as 
\begin{equation}
\Pi_p(z)=\frac{1}{p!}\frac{\partial^p}{\partial \rho^p}\Pi(z,\rho)\bigg|_{\rho=0},\ \ \
\Pi(z,\rho)\equiv
\sum_{d=0}^{\infty}\frac{\Gamma\big((m+2)(d+\rho)+1\big)}{\Gamma(d+\rho+1)^{m+2}}z^{d+\rho}.
\label{pf_cl_sol}
\end{equation}
Two of these solutions are expressed as
\begin{align}
\begin{split}
\Pi_0(z) &=\sum_{d=0}^{\infty}\frac{\big((m+2)d\big)!}{(d!)^{m+2}}z^d,\\
\frac{\Pi_1(z)}{\Pi_0(z)} &=\log z+\frac{m+2}{\Pi_0(z)}\sum_{d=1}^{\infty}\frac{\big((m+2)d\big)!}{(d!)^{m+2}}z^d\left[\Psi\big(1+(m+2)d\big)-\Psi(1+d)\right],
\label{per01}
\end{split}
\end{align}
where $\Psi (x)=\frac{d}{dx}\log\Gamma(x)$ is the digamma function. The complexified K\"ahler modulus $t$ of $X_{m+2}$
is determined by the mirror map
\begin{align}
2\pi it = \frac{\Pi_1(z)}{\Pi_0(z)}.
\label{mirror_map_m}
\end{align}

It has been conjectured in \cite{Jinzenji:2011vm} (and later proved in \cite{Popa:2013jsa}) that the disk partition function (\ref{s_pot_gen_m}) is related to the instanton part of the disk one-point function $\left<{\cal O}_{h^{\mathfrak{m}}}\right>_D^{{X}_{m+2}}$ as
\begin{align}
\left<{\cal O}_{h^{\mathfrak{m}}}\right>_D^{{X}_{m+2}}\Big|_{\scriptsize\mbox{instantons}}=
\frac{2}{I_{\mathfrak{m}}(z)}\theta_z\frac{2}{I_{\mathfrak{m}-1}(z)}\theta_z
\cdots
\theta_z\frac{2}{I_{1}(z)}\theta_z\frac{c^{-1}\tau(z)}{I_0(z)}.
\label{one_pt_spotential}
\end{align}
Here $I_p(z)$ are inductively defined by using closed string periods as
\begin{align}
\begin{split}
I_0(z)&=\Pi_0(z),\\
I_p(z)&=\theta_z\frac{1}{I_{p-1}(z)}\theta_z\frac{1}{I_{p-2}(z)}\theta_z\cdots\theta_z\frac{1}{I_1(z)}\theta_z\frac{\Pi_p(z)}{I_0(z)},
\label{I_p_gen}
\end{split}
\end{align}
which satisfy the relations\footnote{It has been shown in \cite{PopaZinger_co} that the genus zero Gromov-Witten invariants can be obtained from the formula for the three-point function on ${\mathrm{\ICP}^1}$ as
\begin{equation}
\big<{\cal O}_{h}{\cal O}_{h^{p}}{\cal O}_{h^{m-p-1}}\big>_{\mathrm{\ICP}^1}^{X_{m+2}}=
(m+2)\frac{I_{p+1}(z(q))}{I_1(z(q))},
\label{2GW}
\end{equation}
where $q=e^{2\pi it}$ and $z(q)$ is the inverse mirror map derived from (\ref{mirror_map_m}). Utilizing this formula and 
a prescription of \cite{Jockers:2012dk}, it is possible to find exact K\"ahler potential of the K\"ahler moduli space for Calabi-Yau fivefolds with one K\"ahler parameter. It would be interesting to obtain general expression for manifolds with multiple K\"ahler parameters as performed in \cite{Honma:2013hma} for Calabi-Yau fourfolds.}
\begin{equation}
I_b(z)=I_{m+1-b}(z),\ \ \ 
b=1,\ldots,m-1.
\end{equation}
With the use of the mirror map (\ref{mirror_map_m}), by expanding the right hand side of (\ref{one_pt_spotential}) around the large radius point $q=e^{2\pi it}=0$, one obtains the degree $d$ open Gromov-Witten invariant $\left<{\cal O}_{h^{\mathfrak{m}}}\right>_{D,d}^{{X}_{m+2}}$
as\footnote{The open Gromov-Witten invariants with even degree are zero. This can be justified by identifying the disk partition function as a normal function whose expression is generically constrained under the shift $t \rightarrow t+1$. See \cite{Walcher:2006rs, Morrison:2007bm} for more details.}
\begin{equation}
\left<{\cal O}_{h^{\mathfrak{m}}}\right>_D^{{X}_{m+2}}\Big|_{\scriptsize\mbox{instantons}}=\sum_{d=1}^{\infty}\left<{\cal O}_{h^{\mathfrak{m}}}\right>_{D,2d-1}^{{X}_{m+2}}q^{d-\frac{1}{2}}.
\label{one_pt_disk_gen}
\end{equation}
For example, the degree one open Gromov-Witten invariant of $X_{m+2}$ is given by
\begin{equation}
\left<{\cal O}_{h^{\mathfrak{m}}}\right>_{D,1}^{{X}_{m+2}}=2(m+2)!!.
\end{equation}
It is worth noting that the disk two-point functions are obtained from the disk one-point functions and the three-point functions on ${\mathrm{\ICP}^1}$ by
\begin{equation}
\big<{\cal O}_{h^p}{\cal O}_{h^{\mathfrak{m}-p+1}}\big>_{D}^{X_{m+2}}=
\prod_{k=1}^{p-1}
\frac{
\big<{\cal O}_{h}{\cal O}_{h^{\mathfrak{m}-p+k}}{\cal O}_{h^{\mathfrak{m}+p-k}}\big>_{\mathrm{\ICP}^1}^{X_{m+2}}}
{\big<{\cal O}_{h}{\cal O}_{h^{k}}{\cal O}_{h^{2\mathfrak{m}-k}}\big>_{\mathrm{\ICP}^1}^{X_{m+2}}}
\times
2\theta_q
\big<{\cal O}_{h^{\mathfrak{m}}}\big>_{D}^{X_{m+2}},
\label{two_pt_disk_formula}
\end{equation}
where $\theta_q= q\partial/\partial q$. By using the results listed in Table \ref{GW_cy7_x9}--\ref{o_GW_cy11_x13} of Appendix \ref{app:localization}, we can explicitly check this formula. We will revisit this formula in Appendix \ref{app:disk_two}.

In \cite{Jinzenji:2011vm}, a multiple covering formula for the disk one-point function in general dimension was found to be
\begin{equation}
\left<{\cal O}_{h^{\mathfrak{m}}}\right>_D^{{X}_{m+2}}\Big|_{\scriptsize\mbox{instantons}}
=\sum_{d,n=1}^{\infty}\frac{(-1)^{(\mathfrak{m}-1)(n-1)}n^{open}_{2d-1}(h^{\mathfrak{m}})}{2n-1}q^{\frac{(2d-1)(2n-1)}{2}},
\label{multi_cov_cym}
\end{equation}
where the factor $(-1)^{(\mathfrak{m}-1)(n-1)}$ is required to take $n^{open}_{2d-1}(h^{\mathfrak{m}})$ a positive integer. For the case of Calabi-Yau threefolds, by using the divisor equation 
$\left<{\cal O}_{h}\right>_{D,2d-1}^{{X}_{5}}=(2d-1)\left<\emptyset\right>_{D,2d-1}^{{X}_{5}}$ (see, for example, Section 26.3 in \cite{Hori:2003ic}), we find
\begin{align}
\left<{\cal O}_{h}\right>_D^{{X}_{5}}\Big|_{\scriptsize\mbox{instantons}}=
2\theta_q\left<\emptyset\right>_D^{{X}_{5}}\Big|_{\scriptsize\mbox{instantons}},\ \ \ \
\left<\emptyset\right>_D^{{X}_{5}}\Big|_{\scriptsize\mbox{instantons}}\equiv\sum_{d=1}^{\infty}\left<\emptyset\right>_{D,2d-1}^{{X}_{5}}q^{d-\frac{1}{2}}.
\label{divisor_disk_c}
\end{align}
Then, the above expression (\ref{multi_cov_cym}) certainly realize the well-known multiple covering formula for the real quintic \cite{Walcher:2006rs, Pandharipande:2006d}
\begin{equation}
\left<\emptyset\right>_D^{{X}_{5}}\Big|_{\scriptsize\mbox{instantons}}
=\sum_{d,n=1}^{\infty}\frac{n^{open}_{2d-1}}{(2n-1)^2}q^{\frac{(2d-1)(2n-1)}{2}}.
\label{multi_cov_cy3}
\end{equation}
Here $n^{open}_{2d-1} \equiv n^{open}_{2d-1}(h)/(2d-1)$ is a positive integer topological invariant which is called Ooguri-Vafa invariant \cite{Ooguri:1999bv}.

\section{Off-shell formalism via relative periods}\label{sec:disk_GD}

The geometric structure of open/closed B-model and the associated off-shell superpotential depending on the open and closed string moduli has been studied from the viewpoint of ${\cal{N}}=1$ special geometry \cite{Mayr:2001xk, Lerche:2001cw} for the case of non-compact Calabi-Yau threefolds \cite{Lerche:2002ck, Lerche:2002yw}. In \cite{Jockers:2008pe} (see also \cite{Jockers:2009mn}), the study of off-shell superpotential was extended to compact Calabi-Yau threefolds and the corresponding open/closed Picard-Fuchs equation was clarified.

In this section, we utilize the method of \cite{Jockers:2008pe} in general dimensions and explicitly compute the open/closed Picard-Fuchs equation for septic Calabi-Yau fivefold $X_7$. Then we will solve the resulting differential system in the vicinity of the Landau-Ginzburg point, and also check the consistency with the on-shell result.

\subsection{Open string moduli and relative cohomology}\label{subsec:disk_off_shell}

Let us introduce a $2\mathfrak{m}$-dimensional curve $C_u$ depending on an open string modulus $u$ in the mirror Calabi-Yau $Y_{m+2}$. This curve is not necessarily holomorphic, except at the value of $u$ where the on-shell quantity is realized. Consider a chain integral 
\begin{align}
\mathcal{T}_{C_u}(z,u)=\int_{\Gamma_u (u)}\Omega (z),
\label{off_dom_ten}
\end{align}
where $\Gamma_u (u)$ is an $m$-chain whose boundary is defined by $\partial \Gamma_u=C_u$. This is a higher dimensional analog of the off-shell superpotential defined in \cite{Lerche:2002ck, Lerche:2002yw,Jockers:2008pe} 
for Calabi-Yau threefolds.
Here the curve $C_u (u)$ is defined to yield the two holomorphic curves $C_{\pm}$ in (\ref{hol_2m_cyc})
at the critical points of (\ref{off_dom_ten}) with respect to the variation of open string modulus. 

In order to study the moduli dependence of the chain integral (\ref{off_dom_ten}) systematically, it is convenient to introduce the notion of the relative cohomology group and the variation of the mixed Hodge structure \cite{Lerche:2002ck, Lerche:2002yw, Jockers:2008pe}. In what follows, we will briefly look at mathematical preliminaries.

Let us consider the divisor $V$ embedded in the mirror Calabi-Yau manifold $Y_{m+2}$ through the map $i : V \hookrightarrow Y_{m+2}$.\footnote{In order to avoid the difficulties associated with non-holomorphic property of $C_u$, this curve was replaced by the holomorphic divisor $V$ capturing the deformation of $C_u$ in \cite{Lerche:2002ck, Lerche:2002yw, Jockers:2008pe} for Calabi-Yau threefolds. We adapt this argument also in our higher dimensional case and replace the $2\mathfrak{m}$-dimensional curve $C_u$ with the holomorphic divisor $V \subset Y_{m+2}$.} The space of relative forms $\Omega^* (Y_{m+2},V)$ is subspace of the forms $\Omega^* (Y_{m+2})$ induced by the morphism $i^* : \Omega^* (Y_{m+2}) \rightarrow \Omega^* (V)$. Then we can define the relative middle cohomology $H^m(Y_{m+2},V)$ from the complex of pairs $\Omega^p (Y_{m+2}) \oplus \Omega^{p-1} (V)^{-1}$ with differential $d=(d_{Y_{m+2}}+i^*,-d_V)$ as
\begin{align}
\textrm{ker}[H^m(Y_{m+2}) \rightarrow H^m(V)] \oplus 
\textrm{coker}[H^{m-1}(Y_{m+2}) \rightarrow H^{m-1}(V)].
\end{align}
Since $H^m(V)$ is trivial, the first factor is equal to $H^m(Y_{m+2})$ and thus describes the closed string sector. The second factor is called variable cohomology $H^{m-1}_{\rm var}(V)$ of $V$ \cite{Lerche:2002yw}, which varies with the embedding $i : V \hookrightarrow Y_{m+2}$.

Correspondingly, one can consider the relative cohomology group $H^m(Y_{m+2},V)$ and construct a relative $m$-form $\underline{\Xi}$ 
\begin{align}
\underline{\Xi} = (\Xi,\xi) \in H^m(Y_{m+2},V),
\end{align}
as a pair of an $m$-form $\Xi \in H^m(Y_{m+2})$ and an $(m-1)$-form $\xi \in H^{m-1}(V)$, such that $i^*\Xi-d\xi=0$. The equivalence relation is given by
\begin{align}
\underline{\Xi} \sim \underline{\Xi}+(d \alpha,i^* \alpha-d\beta),
\end{align}
where $\alpha$ is an $(m-1)$-form on the mirror Calabi-Yau $Y_{m+2}$ and $\beta$ is an $(m-2)$-form on the divisor $V$.
The pairing between a relative $m$-cycle $\underline{\Gamma}^a$ in the relative homology group $H
_m(Y_{m+2},V)$ and a relative $m$-form $\underline{\Xi} \in H^m(Y_{m+2},V)$ is represented by
\begin{align}
\int_{\underline{\Gamma}^a}\underline{\Xi}
=\int_{\underline{\Gamma}^a}\Xi+\int_{\partial \underline{\Gamma}^a}\xi.
\end{align}
As a result, we can define the relative period integrals
\begin{align}
\underline{\Pi}^a(z,u)=\int_{\underline{\Gamma}^a}\underline{\Omega}(z,u), \ \ \ \ \
\underline{\Gamma}^a \in H_m(Y_{m+2},V,{\IZ}),
\label{off_s_po_rel_p}
\end{align}
which can be regarded as a combination of the integral (\ref{off_dom_ten}) and (\ref{cper}).
Here $\underline{\Omega}(z,u)$ is the relative holomorphic $m$-form which will be later constructed as a residue integral. The relative homology basis $\underline{\Gamma}^a \in H_m(Y_{m+2},V)$ captures the $m$-cycles $\Gamma_p$ in (\ref{cper}) and the $m$-chain $\Gamma_u$ in (\ref{off_dom_ten}).

The convenient method to study the moduli dependence of relative periods is the variation of mixed Hodge structure.
Since the standard Hodge decomposition
\begin{align}
H^m =\bigoplus_{p+q=m} H^{p,q} , \ \ \ H^{q,p}=\overline{H^{p,q}},
\end{align}
does not vary holomorphically under the complex structure deformation, 
it is convenient to use the Hodge filtration
\begin{align}
F^p =\bigoplus_{p \leq a \leq m} H^{a,m-a}(Y_{m+2},V),\ \ \ \ \ \ \ p=0,\ldots,m,
\end{align}
in order to analyze the moduli space in the B-model. Note that
\begin{align}
H^{p,m-p}(Y_{m+2},V) \cong F^p /F^{p+1}.
\end{align}
Another ingredient which is required to define the mixed Hodge structure is the finite increasing weight filtration
\begin{equation}
W_m=H^m(Y_{m+2}), \ \ \
W_{m+1}=H^m(Y_{m+2},V),
\end{equation}
satisfying the relation
\begin{equation}
H^{m-1}_{\rm var}(V) \cong W_{m+1}/W_m.
\end{equation}
Thus the filtration $F^p \cap W_m$ describes the Hodge structure associated with the closed string sector.

The (mixed) Hodge structure can be naturally equipped with the Gauss-Manin connection $\nabla$ satisfying the Griffiths transversality such as $\nabla F^p \subset F^{p-1}$. As a consequence of this transversality, by acting $\nabla$ on the relative holomorphic $m$-form, we can derive the open/closed Picard-Fuchs equations governing the relative period integrals.

A schematical picture of the variation of mixed Hodge structure is given by
\begin{equation}
\xymatrix{
  F^m\cap W_m \ar[r]^{\partial_z} \ar[rd]^{\partial_u}& F^{m-1}\cap W_m  \ar[r]^{\partial_z}  \ar[rd]^{\partial_u}
     & \cdots  \ar[r]^{\partial_z} \ar[rd]^{\partial_u} & F^1\cap W_m  \ar[r]^{\partial_z} \ar[rd]^{\partial_u} & F^0\cap W_m \ar[d]^{\partial_u} \\
  & F^{m-1}\cap W_{m+1} \ar[r]^{\partial_z, \partial_u} & F^{m-2} \cap W_{m+1} \ar[r]^{\quad \ \partial_z, \partial_u} & \cdots \ar[r]^{\partial_z, \partial_u\quad \quad} & F^0 \cap W_{m+1},}
\label{mixedH}
\end{equation}
where $\partial_z$ and $\partial_u$ denote the infinitesimal closed/open string deformations on the space of the relative middle cohomology $H^m(Y_{m+2},V)$, respectively.
As we will see later, each constituent of the above structure can be expressed as a residue integral.

\subsection{The extended Griffith-Dwork algorithm}\label{subsec:offGD}

The Griffiths-Dwork algorithm \cite{griffiths1} is to represent a holomorphic $m$-form on a hypersurface as a residue of a meromorphic $(m+1)$-form on a projective space. In \cite{Jockers:2008pe}, this algorithm was extended to derive the Picard-Fuchs equation governing the relative period integrals of compact Calabi-Yau threefolds. Here we consider the case of higher dimensional compact Calabi-Yau manifolds $Y_{m+2}$.

Let us first define the holomorphic divisor $V \subset Y_{m+2}$ as
\begin{equation}
V: \ \ \ \ Q(\phi) \equiv y_{m+2}^{m+1}-\phi\prod_{i=1}^{m+1}y_i=0,
\label{off_divisor}
\end{equation}
where $\phi$ denotes the open string modulus. Combined with the conditions $y_{2j-1}+y_{2j}=0$, $j=1,\ldots,\mathfrak{m}+1$, this divisor $V$ yields two isolated holomorphic $2\mathfrak{m}$-dimensional curve $C_{\pm}$ in (\ref{hol_2m_cyc}) at the point
\begin{equation}
\phi=(m+2)\psi.
\label{off_divisor_on}
\end{equation}
This means that $(\sqrt{\phi})_{\pm}=\pm \sqrt{(m+2)\psi}$ represent the critical points of the off-shell quantity (\ref{off_dom_ten}), which are supposed to reproduce the correct on-shell result (\ref{h_dom_ten}) in a suitable manner.

Due to the Griffiths transversality, a basis of the relative cohomology group $H^{m}(Y_{m+2},V)$ can be constructed as
\begin{align}
\underline{\pi}& \equiv \left(\pi_{m,m}, \pi_{m-1,m}, \ldots, \pi_{0,m}, \pi_{m-1,m+1}, \pi_{m-2,m+1}, \ldots, \pi_{0,m+1}\right)\nonumber\\
&=
\left(\underline{\Omega}, \partial_\psi\underline{\Omega}, \ldots, \partial_\psi^m\underline{\Omega}, 
\partial_\phi\underline{\Omega}, \partial_\psi\partial_\phi\underline{\Omega}, \ldots, \partial_\psi^{m-1}\partial_\phi\underline{\Omega}\right),
\label{relbase_m}
\end{align}
where the unique relative holomorphic $m$-form $\underline{\Omega}$ in $F^m=H^{m,0}(Y_{m+2},V)$ is defined by a residue integral
\begin{align}
\underline{\Omega}(\psi,\phi) \equiv \int_{\gamma} \frac{\log{Q(\phi)}}{P(\psi)}\Delta,\ \ \ \ \
\Delta \equiv \sum_{i=1}^{m+2} (-1)^i y_i dy_1 \wedge \cdots \wedge \widehat{dy_i} \wedge \cdots \wedge dy_{m+2}.
\label{rel_hol_m_b}
\end{align}
in terms of the polynomials (\ref{cym_hyp_def}) and (\ref{off_divisor}). Here $\widehat{dy_i}$ denotes the exclusion of a component $dy_i$ and the integration is taken over a curve $\gamma$ in the projective space ${\ICP}^{m+1}$ surrounding the hypersurface $Y_{m+2}$. 

The indices of the relative cohomology basis (\ref{relbase_m}) are labeled so that each element corresponds to the constituent of the mixed Hodge structure described in the diagram (\ref{mixedH}). 
By taking the derivatives of $\underline{\Omega}$ with respect to the closed string modulus $\psi$ and open string modulus $\phi$, we obtain the residue integral representations for the closed relative $m$-forms
\begin{align}
\pi_{m-k,m}=(m+2)^{k} k!\int_{\gamma} \frac{\mathrm{u}^{k}}{P(\psi)^{1+k}}\log Q(\phi)\Delta,\ \ \ \ 
k = 0, \ldots, m,
\label{rbasm}
\end{align}
corresponding to the closed string sector and the closed relative $(m-1)$-forms
\begin{align}
\pi_{m-1-\ell,m+1}=-(m+2)^{\ell} \ell! \int_{\hat{\gamma}} \frac{\mathrm{u}^{\ell}\mathrm{v}}{P(\psi)^{1+\ell}Q(\phi)}\Delta,\ \ \ \ 
\ell = 0, \ldots, m-1,
\label{rbasm1}
\end{align}
where $\hat{\gamma}$ is a tube in ${\ICP}^{m+1}$ surrounding the intersection $\{ P=0 \} \cap \{Q=0 \}$.
We have introduced the abbreviations for the monomials as
\begin{equation}
\mathrm{u} \equiv \prod_{i=1}^{m+2}y_i,\ \ \ \ \mathrm{v} \equiv \prod_{i=1}^{m+1}y_i,
\end{equation}
which satisfy the useful formulas
\begin{align}
\mathrm{v}=\frac{Q(\phi)}{\psi-\phi}-\frac{1}{m+2}\frac{\partial_{m+2}P(\psi)}{\psi-\phi},
\label{vQP}
\end{align}
and
\begin{align}
(1-\psi^{m+2})\mathrm{u}^m\mathrm{v}=\sum_{k=1}^{m+2}J_k,
\label{uv_J}
\end{align}
where
\begin{align}
\begin{split}
J_k &\equiv \frac{\psi^{k-1}}{m+2}\Big(\prod_{i=1}^{k-1}y_i^{k-2}\Big)y_k^{k-1}\Big(\prod_{i=k+1}^{m+1}y_i^{m+k}\Big)y_{m+2}^{m+k-1}\partial_{k}P(\psi),\ \ \ k=1,\ldots,m+1, \\
J_{m+2} &\equiv
\frac{\psi^{m+1}}{m+2}\mathrm{u}^{m}\partial_{m+2}P(\psi).
\end{split}
\end{align}

The next step to obtain the Gauss-Manin connection associated with the open/closed Picard-Fuchs operators is to find the derivatives of the basis elements in $\underline{\pi}$ with respect to the open/closed string moduli. Some of the derivatives are, by definition, given by
\begin{align}
\begin{split}
\partial_{\psi}\pi_{k,m}&=\pi_{k-1,m},  \ \ \ \ \ \ \ \ \ \ k=1,2,\ldots,m, \\
\partial_{\phi}\pi_{k,m}&=\pi_{k-1,m+1},  \ \ \ \ \ \ \ k=1,2,\ldots,m, \\
\partial_{\psi}\pi_{k,m+1}&=\pi_{k-1,m+1}, \ \ \ \ \ \ \ k=1,2,\ldots,m-1.
\end{split}
\label{tridebas}
\end{align}

In order to obtain other relations at the level of cohomology, we need to know the residue integral expression of the exact relative forms as well as the closed relative forms. The residue integral expression of an $(m-2)$-form on the divisor $V \subset Y_{m+2}$ is given by
\begin{align}
\beta = \int_{\hat{\gamma}} \sum_{i<j} \frac{y_j q_i (y)-y_i q_j (y)}{P^k Q^{\ell}}(-1)^{i+j} dy_1 \wedge \cdots \wedge \widehat{dy_i} \cdots \wedge \widehat{dy_j} \wedge \cdots \wedge dy_{m+2},
\end{align}
where $q_i(y)$ is a homogeneous polynomial of degree $(k-1)(m+2)+\ell (m+1)+1$. By acting the de Rham differential, one can obtain the exact relative $(m-1)$-form as
\begin{align}
d\beta = \int_{\hat{\gamma}} \sum_{i=1} \left[ k \frac{q_i (y)\partial_i P}{P^{k+1} Q^{\ell}}+\ell \frac{q_i (y)\partial_i Q}{P^{k} Q^{\ell+1}}
-\frac{\partial_i q_i (y)}{P^{k} Q^{\ell}}  \right] \Delta,
\label{exact_beta_k}
\end{align}
where $\partial_i= \partial/\partial y_i$. Similarly one can also construct the exact relative $m$-form on $Y_{m+2}$.

Since the basis elements are now represented by derivatives of the relative holomorphic $m$-form $\underline{\Omega}$, from the linear differential system of $\underline{\pi}$, one can construct the open/closed Picard-Fuchs operators ${\cal{L}}_i$ satisfying
\begin{align}
{\cal{L}}_i \ \underline{\Omega}(\psi,\phi)\simeq 0,
\label{L_i_omega}
\end{align}
which also annihilate the relative periods (\ref{off_s_po_rel_p}). Here the symbol ``$\simeq$'' means that the equation holds at the level of cohomology. 
As we will demonstrate explicitly, in order to construct the open/closed Picard-Fuchs operators for $Y_{m+2}$, it is sufficient to find the expansion of $\partial_{\phi}\pi_{k,m+1}$ and $\partial_{\psi}\pi_{0,m+1}$ in terms of the linear combination of the basis elements $\pi_{k,m+1}$, where $k=0, 1, 2, \ldots, m-1$.

\subsection{Example: Septic Calabi-Yau fivefold}\label{subsec:disk_ex_sep}

As an explicit example, let us consider the septic Calabi-Yau fivefold $X_7$ in the projective space ${\ICP}^6$. 
First we look at the on-shell open Gromov-Witten invariants, which should be encoded in the off-shell formulation described above. Then we will explicitly derive the open/closed Picard-Fuchs equations and solve the system.

\subsubsection{On-shell open Gromov-Witten invariants}\label{subsubsec:septic_on}

The mirror geometry (\ref{cym_hyp_def}) of the septic Calabi-Yau fivefold with a closed string modulus $\psi$ is defined by the homogeneous degree seven polynomial
\begin{align}
Y_7:\ \ P(\psi)=y_1^7+y_2^7+y_3^7+y_4^7+y_5^7+y_6^7+y_7^7-7\psi y_1 y_2 y_3 y_4 y_5 y_6 y_7=0,
\end{align}
in the $\mathbb{Z}^5_7$ orbifold (\ref{Morb}) of the ambient space $\mathbb{CP}^6$. The Hodge diamond\footnote{For general aspects of Calabi-Yau fivefolds such as relation between cohomology groups and the zero modes/fluxes in the M-theory compactification, we refer the reader to \cite{Haupt:2008nu}.} of the mirror septic has the form
\begin{align}
 \begin{array}{ccccccccccc}
   & & & & & 1 & & & & & \\
   & & & & 0 & & 0 & & & & \\
   & & & 0 & & 1667 & & 0 & & & \\ 
   & & 0 & & 0 & & 0 & & 0 & & \\     
   & 0 & & 0 & & 18327 & & 0 & & 0 & \\ 
1 & & 1 & & 1 & & 1 & & 1 & & 1 \\
   & 0 & & 0 & & 18327 & & 0 & & 0 & \\ 
   & & 0 & & 0 & & 0 & & 0 & & \\     
   & & & 0 & & 1667 & & 0 & & & \\ 
   & & & & 0 & & 0 & & & & \\
   & & & & & 1 & & & & &
 \end{array}\label{HD}
\end{align} 

As a mirror of the Lagrangian submanifold $L_7 \cong {\IRP}^5$ in (\ref{inv_brane1}), we consider two isolated holomorphic 4-dimensional curves (\ref{hol_2m_cyc}) in the mirror septic $Y_7$ as
\begin{equation}
C_{\pm}:\ \ 
y_1+y_2=0,\ \ y_3+y_4=0,\ \ y_5+y_6=0,\ \ y_7^3\pm \sqrt{-7\psi}y_1y_3y_5=0.
\end{equation}
As described in (\ref{s_pot_gen_m}), in the vicinity of large complex structure point $z = 1/(7\psi)^7=0$, the chain integral (\ref{h_dom_ten}) associated with the curves $C_{\pm}$ is given by
\begin{align}
\begin{split}
c^{-1}\tau(z)&=2
\sum_{d=1}^{\infty}\frac{\big(7(2d-1)\big)!!}{\big((2d-1)!!\big)^{7}}z^{d-\frac{1}{2}} \\
&=
210z^{\frac12}+\frac{113163050}{9}z^{\frac32}+\frac{64861262508042}{25}z^{\frac52}+\cdots.
\end{split}
\label{tausept}
\end{align}
The closed string periods $\Pi_0(z)$ and $\Pi_1(z)$ are obtained by taking $m=5$ in (\ref{per01}), and the inverse of the mirror map (\ref{mirror_map_m}) is
\begin{align}
z(q) =q-56196 q^2- 3330469926 q^3+ \cdots,
\end{align}
where $q=e^{2\pi it}$. Note that the $\log^2$ solution in (\ref{pf_cl_sol}),
\begin{align}
\begin{split}
\frac{\Pi_2(z)}{\Pi_0(z)} =& \ \frac{1}{2}\log^2 z+(56196 z+ 8067455550 z^2+2236047255069600 z^3) \log z \\
& \ + 144256 z+ 27148468032 z^2+\frac{73519578109500880}{9} z^3 +\cdots,
\end{split}
\end{align}
can be regarded as a generating function of the closed genus zero Gromov-Witten invariants as indicated in (\ref{2GW}),
\begin{align}
\begin{split}
F(q) &\equiv 7 \frac{\Pi_2(z(q))}{\Pi_0(z(q))} \\
&=\frac72\log^2q+\sum_{d=1}^{\infty}\left<{\cal O}_{h^3}\right>_{0,d}^{{X}_{7}}q^d
=\frac72\log^2q+\sum_{d=1}^{\infty}n_d(h^3){\rm Li}_2(q^d),
\end{split}
\label{Fsep}
\end{align}
where the summation in the last equality is referred to as the multiple covering formula \cite{Aspinwall:1991ce, Greene:1993vm}. The list of Gromov-Witten invariants $n_d(h^3)$ are shown in the left of Table \ref{oc_septicGW}.
\begin{table}[t]
\begin{center}
\begin{tabular}{|c|r|}
\hline
$d$ & $n_{d}(h^{3})$ \\ \hline
1 & 1009792 \\
2 & 122239786088 \\
3 & 30528671745480104 \\
4 & 10378199509395886153216 \\ \hline
\end{tabular}
\hspace{4em}
\begin{tabular}{|c|r|}
\hline
$d$ & $n^{open}_{d}(h^{2})$ \\ \hline
1 & 210 \\
3 & 20238540 \\
5 & 7164717071610 \\
7 & 3323817979294765050 \\ \hline
\end{tabular}
\caption{The closed and open Gromov-Witten invariants for the septic Calabi-Yau fivefold $X_7$, which agree with the numbers in the literatures \cite{Greene:1993vm} and \cite{Jinzenji:2011vm}.}
\label{oc_septicGW}
\end{center}
\end{table}

The one-point open Gromov-Witten invariant $\left<{\cal O}_{h^2}\right>_D^{{X}_{7}}$ of septic fivefold is obtained from (\ref{one_pt_spotential}) with $m=5$ (or $\mathfrak{m}=1$)
\begin{equation}
\left<{\cal O}_{h^2}\right>_D^{{X}_{7}}\Big|_{\scriptsize\mbox{instantons}}
=\frac{2}{I_2(z)}\theta_z\frac{2}{I_1(z)}\theta_z\frac{c^{-1}\tau(z)}{\Pi_0(z)}.
\label{1openGW}
\end{equation}
Substituting (\ref{tausept})--(\ref{Fsep}) into (\ref{1openGW}), we finally obtain
\begin{align}
\begin{split}
\left<{\cal O}_{h^2}\right>_D^{{X}_{7}}\Big|_{\scriptsize\mbox{instantons}}
&=
\frac{4\cdot 7}{\theta_q^2F(q)}\theta_q^2\frac{c^{-1}\tau(z(q))}{\Pi_0(z(q))}
\\
&=
210q^{\frac12}+20238470q^{\frac32}+7164717071652q^{\frac52}+\cdots.
\end{split}
\end{align}
where $\theta_q=q\partial/\partial q$ and we have used a relation $\theta_z = I_1(z) \theta_q$. Plugging into the multiple covering formula (\ref{multi_cov_cym}), we obtain the positive integer invariants $n^{open}_{d}(h^{2})$ as listed in the right of Table \ref{oc_septicGW}.

\subsubsection{Off-shell open/closed Picard-Fuchs equations}\label{subsubsec:septic_off}

Having discussed about on-shell properties, let us turn to the application of the off-shell mirror formalism in Section \ref{subsec:offGD} to the septic fivefold. As mentioned in (\ref{off_divisor}), the holomorphic divisor $V \subset Y_7$ depending on an open string modulus $\phi$ is defined by the homogeneous degree six polynomial
\begin{align}
V:\ \ Q(\phi)=y_7^6-\phi y_1y_2y_3y_4y_5y_6=0.
\end{align}

The basis of the relative cohomology group $H^{5}(Y_7,V)$ is defined in (\ref{relbase_m}) as
\begin{align}
\underline{\pi}&=
  \left( \pi_{5,5}, \pi_{4,5}, \pi_{3,5}, \pi_{2,5}, \pi_{1,5}, \pi_{0,5}, \pi_{4,6}, \pi_{3,6}, \pi_{2,6}, \pi_{1,6}, \pi_{0,6} \right) \nonumber \\
&=\left( \underline{\Omega}, \partial_\psi\underline{\Omega}, \partial_\psi^2\underline{\Omega}, \partial_\psi^3\underline{\Omega}, 
\partial_\psi^4\underline{\Omega}, \partial_\psi^5\underline{\Omega},
\partial_\phi\underline{\Omega}, \partial_\psi\partial_\phi\underline{\Omega}, \partial_\psi^2\partial_\phi\underline{\Omega}, \partial_\psi^3\partial_\phi\underline{\Omega}, \partial_\psi^4\partial_\phi\underline{\Omega} \right),
\label{relbase5}
\end{align}
where $\underline{\Omega} (\psi,\phi)$ is the relative holomorphic five-form in (\ref{rel_hol_m_b}) with $m=5$. 
All the basis elements can be represented by the residue integrals as summarized in (\ref{rbasm}) and (\ref{rbasm1}).
In order to derive the open/closed Picard-Fuchs equations governing the relative periods (\ref{off_s_po_rel_p}) for the septic fivefold, we need to construct the linear differential system in terms of the relative cohomology basis (\ref{relbase5}). In the following, we will explicitly describe how to obtain nontrivial derivative relations.

First, by using the aforementioned formula (\ref{vQP}) and the expression for the exact relative forms (\ref{exact_beta_k}) with $m=5$, we can easily find that
\begin{align}
\begin{split}
\partial_{\psi}\partial_{\phi}\underline{\Omega} &\simeq \frac{6\phi}{\psi-\phi} \partial_{\phi}^2\underline{\Omega}+\frac{5}{\psi-\phi} \partial_{\phi}\underline{\Omega}, \\
\partial_{\psi}\partial_{\phi}^2\underline{\Omega} &\simeq  \frac{6\phi}{\psi-\phi} \partial_{\phi}^3\underline{\Omega}+\frac{11\psi-5\phi}{(\psi-\phi)^2} \partial_{\phi}^2\underline{\Omega}+\frac{5}{(\psi-\phi)^2} \partial_{\phi}\underline{\Omega}, \\
\partial_{\psi}^2\partial_{\phi}\underline{\Omega} &\simeq \frac{36\phi^2}{(\psi-\phi)^2} \partial_{\phi}^3\underline{\Omega}+\frac{18\phi(5\psi-3\phi)}{(\psi-\phi)^3} \partial_{\phi}^2\underline{\Omega}+\frac{10(2\psi+\phi)}{(\psi-\phi)^3} \partial_{\phi}\underline{\Omega}.
\end{split}
\end{align}
Similarly, with the use of  (\ref{vQP}) and (\ref{exact_beta_k}), we obtain the relation among higher order derivatives as
\begin{align}
\begin{split}
\partial_{\psi}\partial_{\phi}^3\underline{\Omega} &\simeq \frac{6\phi}{\psi-\phi} \partial_{\phi}^4\underline{\Omega}+\frac{17}{\psi-\phi} \partial_{\phi}^3\underline{\Omega}+\frac{2}{\psi-\phi}\partial_{\psi} \partial_{\phi}^2\underline{\Omega}, \\
\partial_{\psi}^2\partial_{\phi}^2\underline{\Omega} &\simeq  \frac{1}{\psi-\phi} \partial_{\psi}^2\partial_{\phi}\underline{\Omega}+\frac{10}{\psi-\phi} \partial_{\psi}\partial_{\phi}^2\underline{\Omega}+\frac{6\phi}{\psi-\phi} \partial_{\psi}\partial_{\phi}^3\underline{\Omega}, \\
\partial_{\psi}^3\partial_{\phi}\underline{\Omega} &\simeq \frac{6\phi}{\psi-\phi} \partial_{\psi}^2\partial_{\phi}^2\underline{\Omega}+\frac{3}{\psi-\phi} \partial_{\psi}^2\partial_{\phi}\underline{\Omega},
\end{split}
\end{align}
for fourth order,
\begin{align}
\begin{split}
\partial_{\psi}\partial_{\phi}^4\underline{\Omega} &\simeq \frac{3}{\psi-\phi} \partial_{\psi}\partial_{\phi}^3\underline{\Omega}+\frac{23}{\psi-\phi} \partial_{\phi}^4\underline{\Omega}+\frac{6\phi}{\psi-\phi}\partial_{\phi}^5\underline{\Omega}, \\
\partial_{\psi}^2\partial_{\phi}^3\underline{\Omega} &\simeq  \frac{2}{\psi-\phi} \partial_{\psi}^2\partial_{\phi}^2\underline{\Omega}+\frac{16}{\psi-\phi} \partial_{\psi}\partial_{\phi}^3\underline{\Omega}+\frac{6\phi}{\psi-\phi} \partial_{\psi}\partial_{\phi}^4\underline{\Omega}, \\
\partial_{\psi}^3\partial_{\phi}^2\underline{\Omega} &\simeq  \frac{1}{\psi-\phi} \partial_{\psi}^3\partial_{\phi}\underline{\Omega}+\frac{9}{\psi-\phi} \partial_{\psi}^2\partial_{\phi}^2\underline{\Omega}+\frac{6\phi}{\psi-\phi} \partial_{\psi}^2\partial_{\phi}^3\underline{\Omega}, \\
\partial_{\psi}^4\partial_{\phi}\underline{\Omega} &\simeq \frac{2}{\psi-\phi} \partial_{\psi}^3\partial_{\phi}\underline{\Omega}+\frac{6\phi}{\psi-\phi} \partial_{\psi}^3\partial_{\phi}^2\underline{\Omega},
\end{split}
\end{align}
for fifth order and
\begin{align}
\begin{split}
\partial_{\psi}\partial_{\phi}^5\underline{\Omega} &\simeq \frac{4}{\psi-\phi} \partial_{\psi}\partial_{\phi}^4\underline{\Omega}+\frac{29}{\psi-\phi} \partial_{\phi}^5\underline{\Omega}+\frac{6\phi}{\psi-\phi}\partial_{\phi}^6\underline{\Omega}, \\
\partial_{\psi}^2\partial_{\phi}^4\underline{\Omega} &\simeq  \frac{3}{\psi-\phi} \partial_{\psi}^2\partial_{\phi}^3\underline{\Omega}+\frac{22}{\psi-\phi} \partial_{\psi}\partial_{\phi}^4\underline{\Omega}+\frac{6\phi}{\psi-\phi} \partial_{\psi}\partial_{\phi}^5\underline{\Omega}, \\
\partial_{\psi}^3\partial_{\phi}^3\underline{\Omega} &\simeq  \frac{2}{\psi-\phi} \partial_{\psi}^3\partial_{\phi}^2\underline{\Omega}+\frac{15}{\psi-\phi} \partial_{\psi}^2\partial_{\phi}^3\underline{\Omega}+\frac{6\phi}{\psi-\phi} \partial_{\psi}^2\partial_{\phi}^4\underline{\Omega}, \\
\partial_{\psi}^4\partial_{\phi}^2\underline{\Omega} &\simeq  \frac{1}{\psi-\phi} \partial_{\psi}^4\partial_{\phi}\underline{\Omega}+\frac{8}{\psi-\phi} \partial_{\psi}^3\partial_{\phi}^2\underline{\Omega}+\frac{6\phi}{\psi-\phi} \partial_{\psi}^3\partial_{\phi}^3\underline{\Omega}, \\
\partial_{\psi}^5\partial_{\phi}\underline{\Omega} &\simeq \frac{1}{\psi-\phi} \partial_{\psi}^4\partial_{\phi}\underline{\Omega}+\frac{6\phi}{\psi-\phi} \partial_{\psi}^4\partial_{\phi}^2\underline{\Omega},
\end{split}
\end{align}
for sixth order. 

The above relations are not sufficient for the closure of the relevant linear differential system.
To fix the relations completely, we require an additional relation, which can be obtained by (\ref{uv_J}).
After a very long calculation, we obtain a relation for the sixth order derivatives of the relative holomorphic five-form as\begin{align}
&(1-\psi^7)\partial_{\psi}^5\partial_{\phi}\underline{\Omega}\simeq
-\frac{16}{3}\phi(\phi-7\psi)\partial_{\phi}\underline{\Omega}
-\frac{211}{6}\phi^2(\phi-7\psi)\partial_{\phi}^2\underline{\Omega}
-\frac{95}{2}\phi^3(\phi-7\psi)\partial_{\phi}^3\underline{\Omega} \nonumber \\
& \ \ \ \ -\frac{125}{6}\phi^4(\phi-7\psi)\partial_{\phi}^4\underline{\Omega}
-\frac{10}{3}\phi^5(\phi-7\psi)\partial_{\phi}^5\underline{\Omega}
-\frac{1}{6}\phi^6(\phi-7\psi)\partial_{\phi}^6\underline{\Omega} 
-\frac{211}{6}\psi \phi(\phi-7\psi)\partial_{\psi} \partial_{\phi}\underline{\Omega} \nonumber \\
& \ \ \ \ -95\psi \phi^2 (\phi-7\psi)\partial_{\psi} \partial_{\phi}^2 \underline{\Omega}
-\frac{125}{2}\psi \phi^3 (\phi-7\psi)\partial_{\psi} \partial_{\phi}^3 \underline{\Omega}
-\frac{40}{3}\psi \phi^4 (\phi-7\psi)\partial_{\psi} \partial_{\phi}^4 \underline{\Omega} \nonumber \\
& \ \ \ \ -\frac{5}{6}\psi \phi^5 (\phi-7\psi)\partial_{\psi} \partial_{\phi}^5 \underline{\Omega}
-\frac{95}{2}\psi^2 \phi (\phi-7\psi)\partial_{\psi}^2 \partial_{\phi} \underline{\Omega}
-\frac{125}{2}\psi^2 \phi^2 (\phi-7\psi)\partial_{\psi}^2 \partial_{\phi}^2 \underline{\Omega} \nonumber \\
& \ \ \ \ -20\psi^2 \phi^3 (\phi-7\psi)\partial_{\psi}^2 \partial_{\phi}^3 \underline{\Omega}
-\frac{5}{3}\psi^2 \phi^4 (\phi-7\psi)\partial_{\psi}^2 \partial_{\phi}^4 \underline{\Omega}
-\frac{125}{6}\psi^3 \phi (\phi-7\psi)\partial_{\psi}^3 \partial_{\phi} \underline{\Omega} \nonumber \\
& \ \ \ \ -\frac{40}{3}\psi^3 \phi^2 (\phi-7\psi)\partial_{\psi}^3 \partial_{\phi}^2 \underline{\Omega}
-\frac{5}{3}\psi^3 \phi^3 (\phi-7\psi)\partial_{\psi}^3 \partial_{\phi}^3 \underline{\Omega}
-\frac{1}{3}\psi^4 (10\phi (\phi-7\psi)+3\psi^2)\partial_{\psi}^4 \partial_{\phi} \underline{\Omega} \nonumber \\
& \ \ \ \ -\frac{1}{6}\psi^4 \phi (5\phi (\phi-7\psi)+36\psi^2)\partial_{\psi}^4 \partial_{\phi}^2 \underline{\Omega}
-\frac{1}{6}\psi^5 \phi (\phi-\psi)\partial_{\psi}^5 \partial_{\phi} \underline{\Omega}.
\label{deriv_omega_cl_m5}
\end{align}
where we have used (\ref{vQP}) and (\ref{exact_beta_k}) repeatedly.

As a result, we obtain closed relations for the expansion of $\partial_{\phi}\pi_{*,6}$ and 
$\partial_{\psi}\pi_{0,6}$ in terms of the linear combination of the basis elements $\pi_{*,6}$ as
\begin{align}
\begin{split}
\partial_{\phi}\pi_{4,6}&\simeq\frac{\psi-\phi}{6\phi}\pi_{3,6}-\frac{5}{6\phi}\pi_{4,6},\ \ \ \ \ \partial_{\phi}\pi_{3,6}\simeq\frac{\psi-\phi}{6\phi}\pi_{2,6}-\frac{2}{3\phi}\pi_{3,6}, \\
\partial_{\phi}\pi_{2,6}&\simeq\frac{\psi-\phi}{6\phi}\pi_{1,6}-\frac{1}{2\phi}\pi_{2,6},\ \ \ \ \
\partial_{\phi}\pi_{1,6}\simeq\frac{\psi-\phi}{6\phi}\pi_{0,6}-\frac{1}{3\phi}\pi_{1,6},
\end{split}
\label{sGMCY5}
\end{align}
and
\begin{align}
\begin{split}
\partial_{\phi}\pi_{0,6}&\simeq-\frac{7^5 (\phi-\psi) (\phi-7\psi)}{6A}\pi_{4,6}+\frac{7^4\cdot 31(\phi-\psi) (\phi-7\psi)^2}{6A}\pi_{3,6} \\
& \ \ \ \ -\frac{7^3\cdot 90(\phi-\psi) (\phi-7\psi)^3}{6A}\pi_{2,6}+\frac{7^2\cdot 65(\phi-\psi) (\phi-7\psi)^4}{6A}\pi_{1,6} \\
& \ \ \ \ -\frac16\left[\frac{1}{\phi}+\frac{7!!(\phi-\psi) (\phi-7\psi)^5}{A}\right]\pi_{0,6}, \\
\partial_{\psi}\pi_{0,6}&\simeq\frac{7^5 \phi (\phi-7\psi)}{A}\pi_{4,6}-\frac{7^4\cdot 31\phi (\phi-7\psi)^2}{A}\pi_{3,6}
+\frac{7^3\cdot 90\phi (\phi-7\psi)^3}{A}\pi_{2,6} \\
& \ \ \ \ -\frac{7^2\cdot 65\phi (\phi-7\psi)^4}{A}\pi_{1,6}+\frac{7!!\phi (\phi-7\psi)^5}{A}\pi_{0,6},
\end{split}
\label{GMCY5}
\end{align}
where
\begin{align}
A=\phi (\phi-7\psi)^6-6^6.
\end{align}
From the closed derivative relations (\ref{sGMCY5}) and (\ref{GMCY5}), now we can extract the associated open/closed Picard-Fuchs operators which satisfy (\ref{L_i_omega}) and annihilate the relative periods as
\begin{align}
{\cal{L}}_1=\widetilde{\cal{L}}_1 \partial_{\phi}, \ \ \ \  {\cal{L}}_2=\widetilde{\cal{L}}_2 \partial_{\phi},
\label{op_cl_cy5_pf_op}
\end{align}
where $\widetilde{\cal{L}}_1$ and $\widetilde{\cal{L}}_2$ are given by
\begin{align}
\widetilde{\cal{L}}_1&=(\psi-\phi)\psi\partial_{\psi}-6\psi \phi\partial_{\phi}-5\psi,
\label{pf_op_L1}
\\
\widetilde{\cal{L}}_2&=\partial^4_{\psi}\partial_{\phi}+\frac16\left[\frac{1}{\phi}+\frac{7!!(\phi-\psi) (\phi-7\psi)^5}{A}\right]
\partial^4_{\psi}-\frac{7^2\cdot 65(\phi-\psi) (\phi-7\psi)^4}{6A}\partial^3_{\psi} \nonumber \\
& \ \ \ \ +\frac{7^3\cdot 90(\phi-\psi) (\phi-7\psi)^3}{6A}\partial^2_{\psi}-\frac{7^4\cdot 31(\phi-\psi) (\phi-7\psi)^2}{6A}\partial_{\psi}+\frac{7^5(\phi-\psi) (\phi-7\psi)}{6A}.
\label{pf_op_L2}
\end{align}

\subsubsection{Solutions to the off-shell open/closed Picard-Fuchs equations}\label{subsubsec:septic_off_sol}

Now that we have derived the Picard-Fuchs equations capturing the open string deformation, in the remaining part of this section, we will solve the off-shell Picard-Fuchs equations (\ref{op_cl_cy5_pf_op}) in the vicinity of the Landau-Ginzburg point $\psi=0$. As performed in \cite{Jockers:2008pe}, we also check that the inhomogeneous Picard-Fuchs equation (\ref{on_inhom_PF}) can be reproduced at the on-shell point (\ref{off_divisor_on}).

In order to find the relative period integrals annihilated by ${\cal{L}}_1$ and ${\cal{L}}_2$, we first specify the solutions determined by the factorized operators $\widetilde{\cal{L}}_1$ and $\widetilde{\cal{L}}_2$. Let us consider a function
\begin{align}
\chi (\psi, \phi) = \phi^{-\frac{1}{3}}(7\psi-\phi)^3 \lambda (u),
\label{facan}
\end{align}
where $u \equiv \phi (\phi-7\psi)^6$. This function indeed satisfies $\widetilde{\cal{L}}_1 \chi =0$. By acting the operator $\widetilde{\cal{L}}_2$ on this ansatz, we obtain a differential equation for the undetermined function $\lambda (u)$ as
\begin{align}
\left[ \theta_u \left( \theta_u-\frac{1}{6} \right)\left( \theta_u+\frac{1}{2} \right)\left( \theta_u+\frac{1}{3} \right)
\left( \theta_u+\frac{1}{6} \right)-\frac{u}{6^6}\left( \theta_u+\frac{2}{3} \right)^5 \right] \lambda (u)=0.
\end{align}
We can easily find the solutions of this equation as
\begin{equation}
\lambda^{6+\ell}(u)=\frac{1}{\ell!}\; u^{\frac{\ell-3}{6}}\; _5F_4 
\begin{bmatrix}
 \frac{\ell+1}{6}, \frac{\ell+1}{6}, \frac{\ell+1}{6}, \frac{\ell+1}{6}, \frac{\ell+1}{6}; \frac{u}{6^6} \\
 \underbrace{{\textstyle\frac{\ell+2}{6}, \frac{\ell+3}{6}, \frac{\ell+4}{6}, \frac{\ell+5}{6}, \frac{\ell+6}{6}}}
\end{bmatrix},\ \ \ \
\ell=0, \ldots, 4,
\label{lamso}
\end{equation}
where the underbrace means that the unity is to be omitted. Their power series expansions are given by
\begin{align}
\lambda^{6+\ell}(u)=\frac{1}{\Gamma \left( \frac{1+\ell}{6} \right)^6}
\sum_{k=0}^{\infty}\frac{\Gamma (k+\frac{1+\ell}{6})^6}{\Gamma(6k+\ell+1)}u^{k+\frac{\ell-3}{6}}.
\end{align}

Next step to obtain the relative periods determined by ${\cal{L}}_1$ and ${\cal{L}}_2$ is to perform an integration of the solutions (\ref{lamso}) with respect to $\phi$ as represented by
\begin{align}
\underline{\Pi}^{6+\ell}(\psi,\phi)=\int_0^{\phi} \zeta^{-\frac{1}{3}}(7\psi-\zeta)^3 \lambda^{6+\ell} (\zeta (7\psi-\zeta)^6)
d \zeta.
\end{align}
Evaluating these integrals, we finally obtain the relative periods involved with open string deformation as five independent solutions to the equations 
(\ref{op_cl_cy5_pf_op}) as
\begin{align}
\underline{\Pi}^{6+\ell}(\psi,\phi)=\frac{6}{\Gamma (\frac{1+\ell}{6})^6}
\sum_{k=0}^{\infty}\sum_{n=k}^{7k+\ell}\frac{(-1)^{n-k+\ell+1}\Gamma (k+\frac{1+\ell}{6})^6}{(6n+1+\ell)(7k-n+\ell)!(n-k)!}
\phi^{n+\frac{1+\ell}{6}}(7 \psi)^{7k-n+\ell}.
\end{align}
Here we have labeled the relative periods associated with the closed string sector around the Landau-Ginzburg point as
\begin{align}
\underline{\Pi}^{p}(\psi)=\frac{\psi^p}{p!} \; _6F_5 \begin{bmatrix}
 \frac{p+1}{7}, \frac{p+1}{7}, \frac{p+1}{7}, \frac{p+1}{7}, \frac{p+1}{7}, \frac{p+1}{7}; \psi^7 \\
 \underbrace{{\textstyle\frac{p+2}{7}, \frac{p+3}{7}, \frac{p+4}{7}, \frac{p+5}{7}, \frac{p+6}{7}, \frac{p+7}{7}}}
\end{bmatrix}, \ \ \ \ p=0,1,\ldots, 5.
\end{align}

As well as the closed string mirror map $t(\psi) \equiv \underline{\Pi}^1 (\psi) / \underline{\Pi}^0 (\psi)$ around the
Landau-Ginzburg point, now we can also extract the open mirror map by
\begin{align}
\hat{t}(\psi,\phi) \equiv \frac{\underline{\Pi}^6 (\psi)}{\underline{\Pi}^0 (\psi)},
\end{align}
up to an overall constant. The inverse of these mirror maps are 
\begin{align}
\psi (t) &= t-\frac{1}{336}t^8-\frac{20305273}{54486432000}t^{15}+\cdots, \\
\phi^{\frac{1}{6}} (t,\hat{t})
&=-\frac{1}{6}\hat{t}-\frac{1}{2^5 3^3 \cdot 35}t^7\hat{t}+\frac{7^5}{2^{17}3^{15} \cdot 5}t^6\hat{t}^7-\frac{7^5}{2^{22}3^{20} \cdot 65}t^5\hat{t}^{13}+\frac{7^4}{2^{29} 3^{26} \cdot 19}t^4\hat{t}^{19} 
\nonumber\\
& \ \ \ \ -\frac{7^3}{2^{33}3^{33} \cdot 25}t^3\hat{t}^{25}+\cdots.
\end{align}
Then, the higher dimensional analogue of the orbifold disk partition function $W(t,\hat{t})$ in \cite{Jockers:2008pe} is given by
\begin{align}
W(t,\hat{t}) &\equiv \frac{\underline{\Pi}^8 (\psi (t),\phi (t,\hat{t}))}{\underline{\Pi}^0 (\psi (t))} 
\nonumber\\
&= \frac{7}{2^83^{10}}t\hat{t}^9-\frac{1}{2^{15}3^{15} \cdot 5}\hat{t}^{15}-\frac{7^2}{2^33^3}t^2\hat{t}^3
-\frac{7 \cdot 56813}{2^{31}3^{28} \cdot 1235}t^6\hat{t}^{21}+\frac{70290137}{2^{24}3^{22} \cdot 65}t^7\hat{t}^{15} 
\nonumber\\
& \ \ \ -\frac{7 \cdot 1820089}{2^{21}3^{16}}t^8\hat{t}^{9}+\cdots,
\end{align}
whose coefficients can be regarded as the orbifold disk invariants up to an overall normalization constant. It would be 
interesting to check whether these numbers can be reproduced from other independent calculations.

Moreover, from the solution $\underline{\Pi}^8 (\psi,\phi)$, we can extract the on-shell disk partition function around the 
Landau-Ginzburg point $\psi=0$. By taking a difference of the period $\underline{\Pi}^8 (\psi,\phi)$ at the two critical points, we obtain
\begin{align}
\tau_{\textrm{LG}} (\psi)&=\underline{\Pi}^8 (\psi,\phi)|_{ \sqrt{\phi} = \sqrt{7\psi}}-\underline{\Pi}^8 (\psi,\phi)|_{\sqrt{\phi} = -\sqrt{7\psi}} \nonumber \\
&=-\frac{1}{7\psi}\frac{2}{\Gamma (\frac{1}{2})^6}\sum_{k=0}^{\infty}\frac{\Gamma \left( k+\frac{1}{2} \right)^7}{\Gamma \left(7k+\frac{7}{2} \right)} (7 \psi)^{7k+\frac{7}{2}},
\end{align}
which satisfies the inhomogeneous Picard-Fuchs equation associated with $C_{+}-C_{-}$ as
\begin{align}
{\cal{L}}' \tau_{\textrm{LG}} \equiv \left[ \theta_{\psi}(\theta_{\psi}-1)(\theta_{\psi}-2)(\theta_{\psi}-3)(\theta_{\psi}-4)(\theta_{\psi}-5)-\psi^7 (\theta_{\psi}+1)^6
\right] \tau_{\textrm{LG}} =\frac{15}{28 \psi}(7\psi)^{\frac{7}{2}}.
\end{align}
Here we have replaced the standard Picard-Fuchs operator ${\cal{L}}$ in (\ref{on_inhom_PF}) into ${\cal{L}}'  \equiv -7^6 {\cal{L}} \psi$ by taking into account the normalization of the relative holomorphic five-form as described in \cite{Candelas:1990rm,Morrison:2007bm}.
Therefore we conclude that the off-shell Picard-Fuchs equations (\ref{op_cl_cy5_pf_op}) certainly reproduce the 
on-shell properties.

\section{Off-shell formalism via generalized hypergeometric system}\label{sec:disk_glsm}

So far we have seen that the open/closed Picard-Fuchs equations for relative periods can be derived from the Griffith-Dwork method. In this section, we will describe more efficient method to obtain the same Picard-Fuchs system utilizing the generalized hypergeometric system and the gauged linear sigma model (GLSM).
In the context of off-shell open mirror symmetry, this approach was initiated in \cite{Mayr:2001xk, Lerche:2001cw, Lerche:2002yw} and has been applied for non-compact Calabi-Yau threefolds with toric branes. 

Analogously, the open/closed Picard-Fuchs equations for compact Calabi-Yau threefolds can be also obtained from similar framework of the toric geometry \cite{Alim:2009rf, Alim:2009bx, Alim:2010za, Alim:2011rp}. For mathematical details, we refer the reader to \cite{Li:2009dz}. Here we will apply this prescription to the higher dimensional 
Calabi-Yau hypersurfaces $X_{m+2}$. For the septic Calabi-Yau fivefold $X_7$, we explicitly demonstrate that a system of differential equations obtained from this approach is indeed equivalent to the Picard-Fuchs system 
derived in the previous section.

\subsection{Generalized hypergeometric system for period integrals}\label{subsec:disk_glsm_GKZ}

As pioneered in \cite{Witten:1993yc}, various Calabi-Yau manifolds can be realized as the IR fixed points of two dimensional ${\cal{N}}=(2,2)$ GLSM. For example, the abelian GLSM corresponding to the Calabi-Yau hypersurfaces $X_{m+2}$ with one K\"ahler modulus contains a chiral superfield $P=\Phi_0$ with $U(1)$ charge $-m-2$ and $m+2$ chiral superfields $\Phi_i$ 
with $U(1)$ charge $1$. These chiral superfields interact with each other through a gauge invariant superpotential $\mathcal{W}=PG(\Phi_i)$, where $G(\Phi_i)$ is a homogeneous degree $m+2$ polynomial. The structure of this model is determined by the $U(1)$ charge vectors of the chiral superfields
\begin{equation}
l=(l_0,l_1,l_2,\ldots,l_{m+2})=(-m-2,1,1\ldots,1).
\label{cym_hyp_charge}
\end{equation}
as the toric data. As an example with multiple K\"ahler moduli, one can also consider more general $m$-dimensional Calabi-Yau hypersurfaces $X$
described by $U(1)^{n-m-1}$ GLSM with a single superpotential $\mathcal{W}=PG(\Phi_i)$ for $n+1$ chiral superfields $P=\Phi_0$ and $\Phi_i$. Similarly, this system is fixed by $n-m-1$ charge vectors
\begin{equation}
l^{\alpha}=(l_0^{\alpha},l_1^{\alpha},l_2^{\alpha},\ldots,l_n^{\alpha}),\ \ \ \ \ \alpha=1,\ldots,n-m-1.
\label{cym_hyp_charge_vec}
\end{equation}
which satisfy the Calabi-Yau conditions $\sum_{i=0}^{n}l_i^{\alpha}=0$.

As shown in \cite{Hori:2000kt}, a Calabi-Yau manifold $Y$ mirror to $X$ can be described by a Landau-Ginzburg theory. In the viewpoint of two dimensional ${\cal{N}}=(2,2)$ supersymmetry algebra, the mirror symmetry exchanges the left-moving supersymmetry generators as
\begin{align}
Q_{-}\ \longleftrightarrow\ \overline{Q}_{-}.
\end{align}
As a result, the lowest components $\phi_i$ of the chiral superfield $\Phi_i$ in the GLSM on $X$ are related to the lowest components $Y_i$ of the twisted chiral superfields in the mirror Landau-Ginzburg theory as
\begin{align}
\mathrm{Re}(\log Y_i)=-|\phi_i|^2
\end{align}
in the sense of radius inversion symmetry known as T-duality. Similarly, the D-term equations of the GLSM on $X$ depending on the complexified K\"ahler moduli are mapped to the constraints
\begin{equation}
\prod_{i=0}^nY_i^{l_i^{\alpha}}=(-1)^{l_0^{\alpha}}z_{\alpha}
\label{LG_D_mapY}
\end{equation}
where parameters $z_{\alpha}$ represent the complex structure moduli of the mirror manifold $Y$. The mirror Landau-Ginzburg theory also has a twisted superpotential $\widetilde{\mathcal{W}}$
\begin{equation}
\widetilde{\mathcal{W}}=\sum_{i=0}^{n}Y_i
\label{LG_spot_hyp}
\end{equation}
subject to the constraints (\ref{LG_D_mapY}), from which the mirror geometry of $Y$ can be specified.
It can be easily shown that the mirror geometry (\ref{cym_hyp_def}) is derived from the Landau-Ginzburg superpotential (\ref{LG_spot_hyp}) with (\ref{LG_D_mapY}) by using (\ref{z_psi_rel}) and the change of variables $Y_i=y_i^{m+2}$.

As a consequence, the period of the holomorphic $m$-form on $Y$ can be constructed from (\ref{LG_spot_hyp}) and the 
resulting function is known to be annihilated by a set of differential operators of the generalized hypergeometric system of the Gel'fand-Kapranov-Zelevinsky (GKZ) type \cite{GKZ, Hosono:1993qy, Hosono:1994ax}
\begin{align}
\begin{split}
\mathcal{D}_{\alpha}&=\widetilde{\mathcal{D}}_{\alpha}a_0^{-1}, \\
\widetilde{\mathcal{D}}_{\alpha} &\equiv
\prod_{l_i^{\alpha}>0}\left(\frac{\partial}{\partial a_i}\right)^{l_i^{\alpha}}-
\prod_{l_i^{\alpha}<0}\left(\frac{\partial}{\partial a_i}\right)^{-l_i^{\alpha}}.
\end{split}
\label{diff_op_gkz}
\end{align}
Here we have introduced the parameters $a_i$ by scaling the coordinates as $Y_i \to a_i Y_i$ such that $\prod_{i=0}^nY_i^{l_i^{\alpha}}=1$. The constraints (\ref{LG_D_mapY}) are thus replaced by
\begin{equation}
\prod_{i=0}^na_i^{l_i^{\alpha}}=(-1)^{l_0^{\alpha}}z_{\alpha}.
\label{LG_D_map}
\end{equation}

In general, the standard GKZ system defined from $\widetilde{\mathcal{D}}_{\alpha}$ contains redundant solutions as well as the periods \cite{Hosono:1993qy}. 
One common method for finding the full set of periods is to normalize $\widetilde{\mathcal{D}}_{\alpha}$ into $\mathcal{D}_{\alpha}$ as defined in (\ref{diff_op_gkz}) and factorize the resulting expression.
For example, the differential operator (\ref{diff_op_gkz}) for the Calabi-Yau hypersurfaces $X_{m+2}$ specified by the charge vector (\ref{cym_hyp_charge}) takes the form
\begin{equation}
\mathcal{D}=\theta_z\bigg[
\theta_z^{m+1}-(m+2)z\prod_{k=1}^{m+1}\big((m+2)\theta_z+k\big)\bigg],
\label{facPF}
\end{equation}
where we ignored an overall constant $1/a_0a_1\cdots a_{m+2}$. Then one finds the Picard-Fuchs operator defined in (\ref{on_inhom_PF}) 
as an irreducible lower order component of the operator (\ref{facPF}).

\subsection{Extended GKZ system for relative period integrals}\label{subsec:disk_glsm_ext_GKZ}

Let us turn to the inclusion of open string deformation in higher dimensional Calabi-Yau manifolds from the viewpoint of toric geometry.
This can be achieved by introducing the toric branes as first demonstrated for non-compact Calabi-Yau threefolds in \cite{Aganagic:2000gs, Aganagic:2001nx}. 
Recall that the $m$-dimensional Calabi-Yau hypersurfaces $X_{m+2}$ can be defined by a single $U(1)$ charge vector in (\ref{cym_hyp_charge}). In this setup, the off-shell open mirror symmetry can be formulated as follows.

First we consider the set of ``enhanced" $U(1)$ charge vectors defined by
\begin{equation}
{l}=(-m-2, \underbrace{1,1\ldots,1}_{m+2};0,0),\ \ \ \ 
\widehat{l}=(-1,\underbrace{0,0,\ldots,0}_{m+1},1;1,-1),
\label{glsm_hypm_brane}
\end{equation}
where the last two entries represent the $U(1)$ charges of newly introduced chiral superfields of the GLSM on $X_{m+2}$, say $\Phi_{m+3}$ and $\Phi_{m+4}$.\footnote{This set of charge vectors can be regarded to describe a non-compact Calabi-Yau ($m+1$)-fold without branes in a similar fashion as argued in \cite{Alim:2009bx}. We will revisit this aspect and its implications in Section \ref{sec:h_open_closed}.} Note that the sum of the charges in each vector must be zero due to the Calabi-Yau condition.
Let $z$ and $\widehat{z}$ be closed and open string moduli associated with charge vectors $l$ and $\widehat{l}$. According to the formula (\ref{LG_D_mapY}), we obtain
\begin{align}
\frac{1}{Y_0^{m+2}}\prod_{i=1}^{m+2}Y_i=(-1)^{m+2}z,
\end{align}
and also
\begin{align}
\frac{Y_{m+2}Y_{m+3}}{Y_0Y_{m+4}}=-\widehat{z},
\end{align}
as an additional constraint.
Here the new coordinates $Y_{m+3}$ and $Y_{m+4}$ are mirror dual to the lowest components of the superfields $\Phi_{m+3}$ and $\Phi_{m+4}$, respectively. 
Taking into account the contributions from $Y_{m+3}$ and $Y_{m+4}$ in the twisted superpotential (\ref{LG_spot_hyp}), we obtain a defining equation of a submanifold
\begin{equation}
y_{m+2}^{m+1}+(m+2)\psi \widehat{z}\prod_{i=1}^{m+1}y_i=0,
\end{equation}
as well as the mirror geometry (\ref{cym_hyp_def}).
This implies that the enhanced 
charge vectors (\ref{glsm_hypm_brane}) indeed capture the holomorphic divisor (\ref{off_divisor}) if we identify the off-shell open string deformation parameter as
\begin{equation}
\phi=-(m+2)\psi \widehat{z}.
\label{hat_z_phi_rel}
\end{equation}
Comparing with (\ref{off_divisor_on}), we find that the on-shell behaviour can be realized by taking
\begin{equation}
\widehat{z}=-1.
\label{zha}
\end{equation}

Now we move on to the derivation of associated hypergeometric system.
By changing the basis of the charge vectors (\ref{glsm_hypm_brane}), we choose
\begin{equation}
l^1 \equiv l-\widehat{l}=(-m-1,\underbrace{1,1,\ldots,1}_{m+1},0;-1,1),\ \ \ \ \ \ 
l^2 \equiv \widehat{l}=(-1,\underbrace{0,0,\ldots,0}_{m+1},1;1,-1),
\label{glsm_hypm_brane_c}
\end{equation}
for later convenience. Correspondingly, the associated moduli parameters are transformed into the form
\begin{equation}
z_1=z \widehat{z}^{-1},\ \ \ \ 
z_2=\widehat{z}
\label{z_1_2_z_hat_z}
\end{equation}
to satisfy the constraint (\ref{LG_D_mapY}).

Substituting the set of $U(1)$ charge vectors (\ref{glsm_hypm_brane_c}) into the general formula of GKZ operators (\ref{diff_op_gkz}), we obtain differential equations of the extended GKZ system of Calabi-Yau hypersurfaces $X_{m+2}$ with a toric brane as
\begin{align}
\begin{split}
&
\mathcal{D}_1=\bigg[\theta_1^{m+1}+z_1\prod_{k=1}^{m+1}\big( (m+1)\theta_1+\theta_2+k\big) \bigg](\theta_1-\theta_2),\\
&
\mathcal{D}_2=\big[\theta_2+z_2\big( (m+1)\theta_1+\theta_2+1\big) \big](\theta_1-\theta_2).
\end{split}
\end{align}
As already mentioned above, this kind of GKZ system can be factorizable. Indeed, we see that
$(m+1)\mathcal{D}_1+\theta_1^m\mathcal{D}_2=\left[ (m+1)\theta_1+\theta_2 \right] \widetilde{\mathcal{D}}_1$, where
\begin{align}
\widetilde{\mathcal{D}}_1\equiv \bigg[(1+z_2)\theta_1^m+
(m+1)z_1\prod_{k=1}^m \big( (m+1)\theta_1+\theta_2+k \big) \bigg](\theta_1-\theta_2).
\label{D1_D2_factor}
\end{align}
Note that in order to include the closed string sector and obtain a complete system of differential equations, 
we also need to consider one more operator $\mathcal{D}_{3}$ associated with the charge vector ${l}=l_1+l_2$ given by
\begin{equation}
\mathcal{D}_{3}=\theta_1^{m+1}\theta_2-
z_1z_2\prod_{k=1}^{m+2}\big( (m+1)\theta_1+\theta_2+k\big).
\end{equation}
Again we can factorize the operator as $(m+2)\mathcal{D}_{3}+\theta_1^m\mathcal{D}_2=\left[ (m+1)\theta_1+\theta_2\right] \widetilde{\mathcal{D}}_{3}$ where
\begin{align}
\widetilde{\mathcal{D}}_{3} \equiv \bigg[\big( \theta_2-z_2(\theta_1-\theta_2)\big) \theta_1^m-
(m+2)z_1z_2\prod_{k=1}^{m+1}\big( (m+1)\theta_1+\theta_2+k\big) \bigg].
\label{D_D2_factor}
\end{align}

Eventually we find that a complete set of differential equations from the extended GKZ system is determined by 
$\{\widetilde{\mathcal{D}}_1, \mathcal{D}_2, \widetilde{\mathcal{D}}_{3}\}$. 
In the following section, we will investigate several aspects of this system and verify that the toric geometry approach for higher dimensional Calabi-Yau hypersurfaces indeed satisfy nontrivial checks.

\subsection{Picard-Fuchs equations from extended GKZ operators}
\label{subsec:PF_ext_GKZ}

\subsubsection{On-shell disk partition function}

Here we will first explicitly show that from the solutions to the extended GKZ system, the on-shell disk partition function (\ref{s_pot_gen_m}) is precisely reproduced at the specific point of moduli space.

Recall that the expression (\ref{s_pot_gen_m}) is expanded around the large complex structure point $z=0$.
According to (\ref{zha}) and (\ref{z_1_2_z_hat_z}), the on-shell solution should be realized from the expansion around
$(z_1, z_2)=(0,-1)$. Therefore, as performed in \cite{Alim:2009rf} for the quintic Calabi-Yau threefold, it is convenient to reparametrize $z_1$ and $z_2$ as
\begin{equation}
u=(-z_1)^{-\frac{1}{m+1}}(1+z_2), \ \ \ \ 
v=(-z_1)^{\frac{1}{m+1}},
\end{equation}
so that the on-shell critical point is represented by $u=0$.
In these variables, we consider a kernel of the extended GKZ operators $\{\widetilde{\mathcal{D}}_1, \mathcal{D}_2, \widetilde{\mathcal{D}}_{3}\}$ satisfying
\begin{equation}
(\theta_1-\theta_2)\tau_{\mbox{\scriptsize off}}(u,v)=\widetilde{\mathcal{D}}_{3}\tau_{\mbox{\scriptsize off}}(u,v)=0.
\end{equation}
Solving these equations, we find that the function $\tau_{\mbox{\scriptsize off}}$ takes the form
\begin{equation}
\tau_{\mbox{\scriptsize off}}(u,v)=\sum_{i,j=0}^{\infty}a_{i,j}u^iv^j,
\end{equation}
where the coefficients $a_{i,j}$ are subject to the following recurrence relations:
\begin{align}
\begin{split}
&
a_{i+1,j+1}=\frac{(m+2)i-j}{(m+1)(i+1)}a_{i,j},\\
&
a_{i,j+m+1}=\frac{\prod_{k=1}^{m+2}\big((m+2)(j-i)+(m+1)k \big)}
{\big(j-(m+2)i+(m+1)\big)(j-i+m+1)^{m+1}}a_{i,j}.
\end{split}
\end{align}

To analyze the on-shell behavior, it is sufficient to look at the terms depending only on the variable $v$.
With the use of the parametrization $m=2\mathfrak{m}+1 \ (\mathfrak{m} \in {\mathbb{N}})$, 
we further obtain the expression for odd dimensional Calabi-Yau hypersurfaces as
\begin{align}
a_{0,(\mathfrak{m}+1)(2d-1)}&=\frac{\prod_{k'=0}^{m+1}\big((m+2)(2d-1)-2k'\big)}
{(2d-1)^{m+2}}a_{0,(\mathfrak{m}+1)(2d-3)}
\nonumber\\
&=
\frac{\big((m+2)(2d-1)\big)!!}{(m+2)!!\big((2d-1)!!\big)^{m+2}}a_{0,\mathfrak{m}+1},
\end{align}
where $d \in{\IN}$.
Taking a normalization such that $a_{0,\mathfrak{m}+1}=2c(m+2)!!$, we see that 
the above solution leads to (\ref{s_pot_gen_m}) as
\begin{equation}
\tau_{\mbox{\scriptsize off}}\big(u=0,v=z^{\frac{1}{m+1}}\big)=\tau(z).
\end{equation}

\subsubsection{Off-shell Picard-Fuchs equations}

Next, we will compare the system $\{\widetilde{\mathcal{D}}_1, \mathcal{D}_2\}$ of differential equations determined by extended GKZ operators 
with our previously obtained open/closed Picard-Fuchs equations in (\ref{op_cl_cy5_pf_op}) for the septic Calabi-Yau fivefold. 

From (\ref{z_psi_rel}), (\ref{hat_z_phi_rel}), and (\ref{z_1_2_z_hat_z}), we can rewrite the variables $z_1$, $z_2$ in terms of $\psi$ and $\phi$ as
\begin{equation}
z_1=-\frac{1}{\big((m+2)\psi\big)^{m+1}\phi},\ \ \ \
z_2=-\frac{\phi}{(m+2)\psi}.
\end{equation}
Then the GKZ operators can be expressed up to overall constant factors as
\begin{align}
&
\widetilde{\mathcal{D}}_1=
\bigg[\big((m+2)\psi-\phi\big)\psi^m\phi (\theta_{\psi}+\theta_{\phi}+2)^m
-(m+1)\prod_{k=1}^m(\theta_{\psi}-i+1)\bigg]\partial_{\phi}\psi^{-1},\\
&
\mathcal{D}_2=
\big[(\psi-\phi)\theta_{\psi}-(m+1)\psi\theta_{\phi}-m\psi\big]
\partial_{\phi}\psi^{-1} \equiv \mathcal{P}_1\partial_{\phi}\psi^{-1},
\label{pf_op_p1}
\end{align}
where $\theta_{\psi}=\psi \partial_{\psi}$ and $\theta_{\phi}=\phi \partial_{\phi}$.
Using (\ref{pf_op_p1}), the operator $\widetilde{\mathcal{D}}_1$ can be further evaluated as
\begin{align}
\mathcal{P}_2\partial_{\phi}\psi^{-1}& \equiv
\bigg[\big((m+2)\psi-\phi\big)\psi^m\phi \left[
\big((m+2)\psi-\phi\big)\partial_{\psi}+m+2\right]^m
\nonumber\\
&\hspace{11em}\;
-(m+1)^{m+1}\prod_{k=1}^m(\theta_{\psi}-i+1)\bigg]\partial_{\phi}\psi^{-1}.
\label{pf_op_p2}
\end{align}

Obviously, in the case of $m=5$, the differential operator $\mathcal{P}_1$ in (\ref{pf_op_p1}) agrees with the 
factorized Picard-Fuchs operator $\widetilde{\cal{L}}_1$ in (\ref{pf_op_L1}). Meanwhile, using the operator ${\cal{L}}_1$ in (\ref{op_cl_cy5_pf_op}), the remaining Picard-Fuchs operator ${\cal{L}}_2$ can be expressed as ${\cal{L}}'_{2} \partial_{\phi}$, where
\begin{align}
{\cal{L}}'_{2}&=
A\partial_{\psi}^5+7!!\phi(7\psi-\phi)^5\partial_{\psi}^4
+7^2\cdot 65\phi(7\psi-\phi)^4\partial_{\psi}^3+7^3\cdot 90\phi(7\psi-\phi)^3\partial_{\psi}^2
\nonumber\\
&\ \ \ \
+7^4\cdot 31\phi(7\psi-\phi)^2\partial_{\psi}+7^5\phi(7\psi-\phi),
\end{align}
up to an overall constant.
Comparing with the operator $\mathcal{P}_2$ in (\ref{pf_op_p2}), we find that
\begin{equation}
\mathcal{P}_2={\cal{L}}'_{2}.
\end{equation}
Therefore it turns out that the kernel of the differential operators $\{\widetilde{\mathcal{D}}_1\psi, \mathcal{D}_2\psi\}$ 
constructed from the toric geometry prescription indeed coincides with that of the open/closed Picard-Fuchs operators (\ref{op_cl_cy5_pf_op}).

\section{Open-closed duality in higher dimensions}\label{sec:h_open_closed}

In the previous section, we discussed the extended GKZ system associated with the charge vectors (\ref{glsm_hypm_brane_c}) which describes the open topological strings on the $m$-dimensional Calabi-Yau hypersurface $X_{m+2}$. Alternatively, we can also consider that these charge vectors describe a non-compact Calabi-Yau ($m+1$)-fold $\widehat{X}_{m+2}$ embedded in the non-compact space without brane. To discuss this description more precisely, as considered in \cite{Alim:2009bx} for $m=3$, let us compactify the embedding space by further adding a chiral superfield $\Phi_{m+5}$ and an extra charge vector $\widehat{l}^3$ as
\begin{align}
&
\widehat{l}^1=(&-m-&1,&&1,&&1,&&\underbrace{\cdots}_{m-2},&&1,&&0;&-&1,&&1,&&0&),
\nonumber\\
&
\widehat{l}^2=(&-&1,&&0,&&0,&&\underbrace{\cdots}_{m-2},&&0,&&1;&&1,&-&1,&&0&),\\
&
\widehat{l}^3=(&&0,&-&2,&&0,&&\underbrace{\cdots}_{m-2},&&0,&&0;&&0,&&1,&&1&),
\nonumber
\end{align}
where the last entry represents the $U(1)$ charges of $\Phi_{m+5}$. 
Then one obtains the compact elliptically fibered Calabi-Yau ($m+1$)-fold $\widehat{X}^c_{m+2}$ whose base ${\ICP}^1$ has the K\"ahler form associated with the extra charge vector $\widehat{l}^3$.

Let $t_i$ and $h_i$, $i=1,2,3$ be the K\"ahler moduli and K\"ahler forms associated with the charge vectors $\widehat{l}^i$. The generating function of the classical intersection numbers $\kappa_{i_1i_2 \ldots i_{m+1}}=\int_{\widehat{X}^c_{m+2}}h_{i_1}\wedge h_{i_2}\wedge \cdots \wedge h_{i_{m+1}}$ are determined by the charge vectors as
\begin{align}
\begin{split}
K_{m+1}&\equiv
\frac{1}{(m+1)!}\sum_{i_1,i_2,\ldots,i_{m+1}=1,2,3}\kappa_{i_1i_2 \ldots i_{m+1}}t_{i_1}t_{i_2}\cdots t_{i_{m+1}}
\\
&=
\frac{2(m+2)}{(m+1)!}t^{m+1}-\frac{m+1}{(m+1)!}t_1^{m+1}
+\frac{m+2}{m!}t^ms,
\end{split}
\end{align}
where we have defined
\begin{equation}
t\equiv t_1+t_2,\ \ \ \
s\equiv t_3,
\end{equation}
and correspondingly
\begin{equation}
h\equiv h_1+h_2,\ \ \ \
S\equiv h_3.
\end{equation}
Here we consider the odd dimensional case with $m=2\mathfrak{m}+1$. 
Taking the volume of the base ${\ICP}^1$ to infinity (${\rm Im}\; s \to \infty$) corresponds to the decompactification limit $\widehat{X}^c_{m+2} \to \widehat{X}_{m+2}$. In this limit, the following $2\mathfrak{m}$ independent two-point functions on ${\ICP}^1$ are relevant:\footnote{In the case of the quintic Calabi-Yau threefold, our $\big<{\cal O}_{h}{\cal O}_{hS}\big>_{\mathrm{\ICP}^1}^{\widehat{X}^c_{5}}$ and 
$\big<{\cal O}_{h_1}{\cal O}_{h_1^{2}}\big>_{\mathrm{\ICP}^1}^{\widehat{X}^c_{5}}$ correspond to $\Pi_{2,2}$ and $\Pi_{2,3}$ in the Appendix A of \cite{Alim:2009bx}, respectively.}
\begin{equation}
\big<{\cal O}_{h^{p}}{\cal O}_{h^{2\mathfrak{m}-p}S}\big>_{\mathrm{\ICP}^1}^{\widehat{X}^c_{m+2}},\ \ \ \ 
\big<{\cal O}_{h_1^{p}}{\cal O}_{h_1^{2\mathfrak{m}-p+1}}\big>_{\mathrm{\ICP}^1}^{\widehat{X}^c_{m+2}},\ \ \ \ 
p=1,\ldots,\mathfrak{m}.
\label{two_point_X_comp}
\end{equation}
These are the generating functions of the degree $d$ genus zero Gromov-Witten invariants $\big<{\cal O}_{h^{p}}{\cal O}_{h^{2\mathfrak{m}-p}S}\big>_{0,d}^{\widehat{X}^c_{m+2}}$ and $\big<{\cal O}_{h_1^{p}}{\cal O}_{h_1^{2\mathfrak{m}-p+1}}\big>_{0,d}^{\widehat{X}^c_{m+2}}$ of $\widehat{X}^c_{m+2}$.

We see that in the limit ${\rm Im}\ s \to \infty$, the first two-point functions in (\ref{two_point_X_comp}) coincide with the two-point functions for the Calabi-Yau $m$-fold $X_{m+2}$:\footnote{For $m=5$, we will check this coincidence in Appendix \ref{app:sol_ext_gkz}.}
\begin{equation}
\big<{\cal O}_{h^{p}}{\cal O}_{h^{2\mathfrak{m}-p}S}\big>_{\mathrm{\ICP}^1}^{\widehat{X}^c_{m+2}}\Big|_{{\rm Im}\; s \to \infty}=
\big<{\cal O}_{h^{p}}{\cal O}_{h^{2\mathfrak{m}-p}}\big>_{\mathrm{\ICP}^1}^{X_{m+2}}.
\label{closed_two_X_comp}
\end{equation}
The remaining two-point functions in (\ref{two_point_X_comp}) are naturally expected to give the following off-shell disk partition functions in the vicinity of the large radius point:
\begin{equation}
\big<{\cal O}_{h_1^{p}}{\cal O}_{h_1^{2\mathfrak{m}-p+1}}\big>_{\mathrm{\ICP}^1}^{\widehat{X}^c_{m+2}}\Big|_{{\rm Im}\; s \to \infty}=
\big<{\cal O}_{h_1^{p}}{\cal O}_{h_1^{2\mathfrak{m}-p+1}}\big>_{\mathrm{\ICP}^1}^{\widehat{X}_{m+2}}\equiv
\big<{\cal O}_{h_1^{p}}\big>_{D_p}^{X_{m+2}}.
\end{equation}
The last equality means that the marked point on $\mathrm{\ICP}^1$ for the operator ${\cal O}_{h_1^{2\mathfrak{m}-p+1}}$ transformed into a boundary (or hole) on which the disk $D_p$ can end. In other words, the boundary $\partial D_p$ of the disk $D_p$ should be mapped to a real ($2p+1$)-dimensional subspace of the off-shell Lagrangian submanifold in $X_{m+2}$. This relation is a higher dimensional generalization of the open-closed string duality for Calabi-Yau threefolds discussed in \cite{Mayr:2001xk, Lerche:2001cw, Alim:2009bx, Aganagic:2009jq}. It would be interesting to check this relation by explicitly defining and computing the off-shell disk partition functions in the A-model. Note that even for Calabi-Yau threefolds, this problem has not been fully elucidated.

\section{Conclusions and discussions}
\label{sec:conclusion}

We studied the open and closed string deformations for a class of Calabi-Yau hypersurfaces in general dimensions.
To formulate the open mirror symmetry and extract the generating functions for  topological invariants, we employed two methods using $1)$ the Gauss-Manin system for the relative cohomology group and $2)$ the GKZ system associated with the generalized hypergeometric functions. We have shown that these independent approaches yield the same differential system even for higher dimensions.
In particular, for the simplest case of the compact Calabi-Yau fivefold, we have explicitly derived the Picard-Fuchs system depending on the open and closed string moduli, and also evaluated the off-shell open Gromov-Witten invariants in the Landau-Ginzburg point and the large radius point. Moreover, to justify our present construction of the off-shell higher dimensional open mirror symmetry, we have confirmed that the on-shell properties associated with the involution brane can be correctly reproduced at the specific values of the open string deformation parameter.

Throughout this paper, we have paid attention to the compact Calabi-Yau hypersurfaces. Meanwhile, as explored in \cite{Mayr:2001xk, Lerche:2001cw, Lerche:2002yw}, it can be possible to formulate the off-shell mirror symmetry for the noncompact Calabi-Yau in a similar way.
To generalize our higher dimensional construction into the local Calabi-Yau $m$-folds may shed light on novel aspects of the local mirror symmetry \cite{Chiang:1999tz} and the geometric engineering \cite{Katz:1996fh}.

From the viewpoint of flux compactifications, the closed string periods are related to the Gukov-Vafa-Witten type flux superpotential \cite{Gukov:1999ya}, and the relative periods are correspond to the superpotential arising from branes wrapping the internal cycles of the manifold. Although it is not obvious whether our result for $m>5$ can be useful for the space-time compactification, there exists a possible application for the case of $m=5$, namely the Calabi-Yau fivefolds. Indeed, in \cite{Haupt:2008nu}, the flux compactification of M-theory on Calabi-Yau fivefolds and its relation to the one-dimensional super-mechanics theories has been intensely studied. Naively we expect that holomorphic disks in Calabi-Yau fivefolds considered in this work are related to the superpotential in the $2a$ sector of this theory. It would be quite interesting to contemplate whether one can apply higher dimensional mirror prescription into this fascinating setup.

Further study on the open-closed duality is also important. Although we have shown that the charge vectors for brane geometry defined in the off-shell mirror formalism are naturally interpreted to describe another manifold without branes,
a more detailed analysis is required to understand the correspondence completely. It would be necessary to reinterpret the off-shell mirror symmetry in the A-model viewpoint for the brane geometry and find the whole structure of the closed Gromov-Witten partition functions for higher dimensional Calabi-Yau manifolds with multiple K\"ahler parameters for the closed dual geometry.

Finally, we comment that there is another interesting direction for the future research about higher dimensional Calabi-Yau manifolds. It has been clarified that the Type IIB superstring on $AdS_5 \times S^5$ admits an A-model formulation through the pure spinor formalism and it was shown in \cite{Berkovits:2007rj} that the Coulomb branch of this model can be captured by a GLSM on a superprojective space. 
Thus one can expect that to analyze the A-model on the specific Calabi-Yau supermanifolds may provide some insights into the AdS/CFT correspondence from the worldsheet perspective. Since this particular type of supermanifolds 
are supposed to be related to the (typically higher dimensional) local Calabi-Yau varieties by means of reversion of Grassmann parity of the fibers \cite{Schwarz:1995ak}, it would be quite useful to establish the higher dimensional local mirror symmetry by modifying our present construction and the works in \cite{Mayr:2001xk, Lerche:2001cw, Lerche:2002yw}.

\subsection*{Acknowledgements}

We would like to thank Hiroyuki Fuji, Albrecht Klemm, Satoshi Nawata, Masoud Soroush, Piotr Su{\l}kowski, and Yoske Sumitomo for useful discussions and comments. We are grateful to KEK Theory Workshop 2015, and stringtheory.pl annual meeting 2015 for the opportunity to present some preliminary results of this work. We are grateful to Harish-Chandra Research Institute where this project was initiated. MM gratefully acknowledges support from the Simons Center for Geometry and Physics, Stony Brook University at which some of the research for this paper was performed. The work of YH is supported by the ERC Starting Grant QC \& C no. 259562 ``Quantum fields and Curvature--Novel Constructive Approach via Operator Product Expansion''. The work of MM is supported by the ERC Starting Grant FIELDS-KNOTS no. 335739 ``Quantum fields and knot homologies''.

\appendix
\section{Direct Calculation in the A-model via localization}\label{app:localization}

In this appendix, we describe how to compute the closed and open Gromov-Witten invariants of higher dimensional Calabi-Yau hypersurfaces in the A-model via the localization \cite{Kontsevich:1994na}. The direct enumeration of the holomorphic spheres and disks can be done by using the localization.

\subsection{Counting holomorphic spheres}\label{subapp:localization_close}

Here we consider the Calabi-Yau $m$-fold $X_{m+2} \subset {\ICP}^{m+1}$ defined by a degree $m+2$ hypersurface in the A-model. Let ${\cal O}_{h^i}$ be observables defined by the elements $h^i \in H^{2i}({\ICP}^{m+1},{\IC})$, where $h \in H^2({\ICP}^{m+1},{\IC})$ is the hyperplane class of ${\ICP}^{m+1}$. The genus $0$ and degree $d$ Gromov-Witten invariant is defined by
\begin{equation}
\left<{\cal O}_{h^{p_1}}\cdots{\cal O}_{h^{p_n}}\right>_{0,d}^{{X}_{m+2}} \equiv
\int_{\overline{\cal M}_{0,n}({\ICP}^{m+1},d)}c_{(m+2)d+1}({\cal E}_d)ev^*_1(h^{p_1})\cdots ev^*_n(h^{p_n}),
\label{closed_GWA}
\end{equation}
where $\overline{\cal M}_{0,n}({\ICP}^{m+1},d)$ is the compactified moduli space of degree $d$ and 
$n$-pointed stable maps from genus $0$ stable curves to ${\ICP}^{m+1}$. The map $ev_i$ is the evaluation map at the $i$-th marked point from the moduli space $\overline{\cal M}_{0,n}({\ICP}^{m+1},d)$ to ${\ICP}^{m+1}$. Here ${\cal E}_d=R^0\pi_*^{n+1} ev^*_{n+1}{\cal O}_{{\ICP}^{m+1}}(m+2)$ denotes the coherent sheaf on $\overline{\cal M}_{0,n}({\ICP}^{m+1},d)$ associated with the direct image $R^0$ under the forgetful map $\pi^{n+1}$: $\overline{\cal M}_{0,n+1}({\ICP}^{m+1},d) \to \overline{\cal M}_{0,n}({\ICP}^{m+1},d)$ of the pullback of ${\cal O}_{{\ICP}^{m+1}}(m+2)$ by the evaluation map at the $(n+1)$-th marked point. Note that, in order to obtain the nontrivial quantity in (\ref{closed_GWA}), the constraint\footnote{The complex dimension of the moduli space is $\dim_{{\IC}}\overline{\cal M}_{0,n}({\ICP}^{m+1},d)=(m+2)d+m-2+n$, and the rank of coherent sheaf ${\cal E}_d$ is $(m+2)d+1$.}
\begin{equation}
\sum_{i=1}^np_i=m+n-3,
\label{select_rule}
\end{equation}
should be satisfied.

\begin{figure}[t]
 \centering
  \includegraphics[width=125mm]{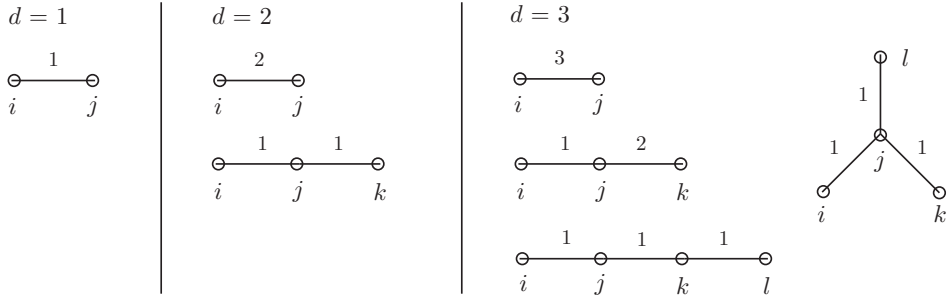}
 \caption{Tree graphs without marked points up to degree $d=3$}
\label{closed_1}
\end{figure}

\begin{figure}[t]
 \centering
  \includegraphics[width=125mm]{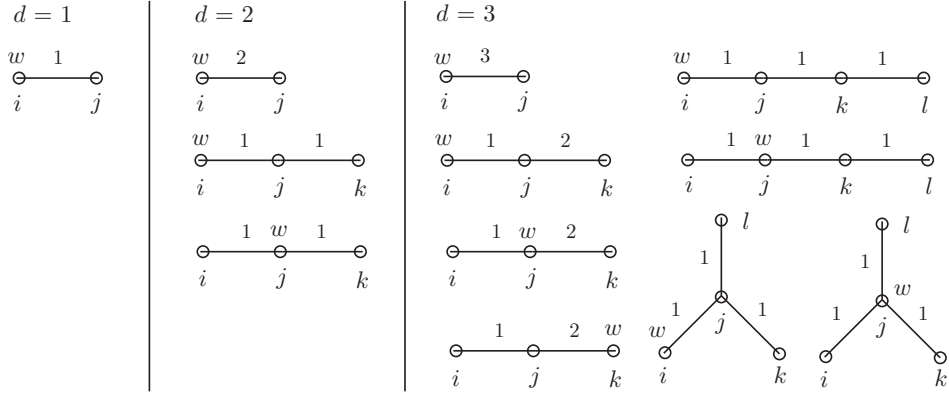}
 \caption{Tree graphs with one marked point $w$ up to degree $d=3$}
\label{closed_2}
\end{figure}

\begin{figure}[t]
 \centering
  \includegraphics[width=125mm]{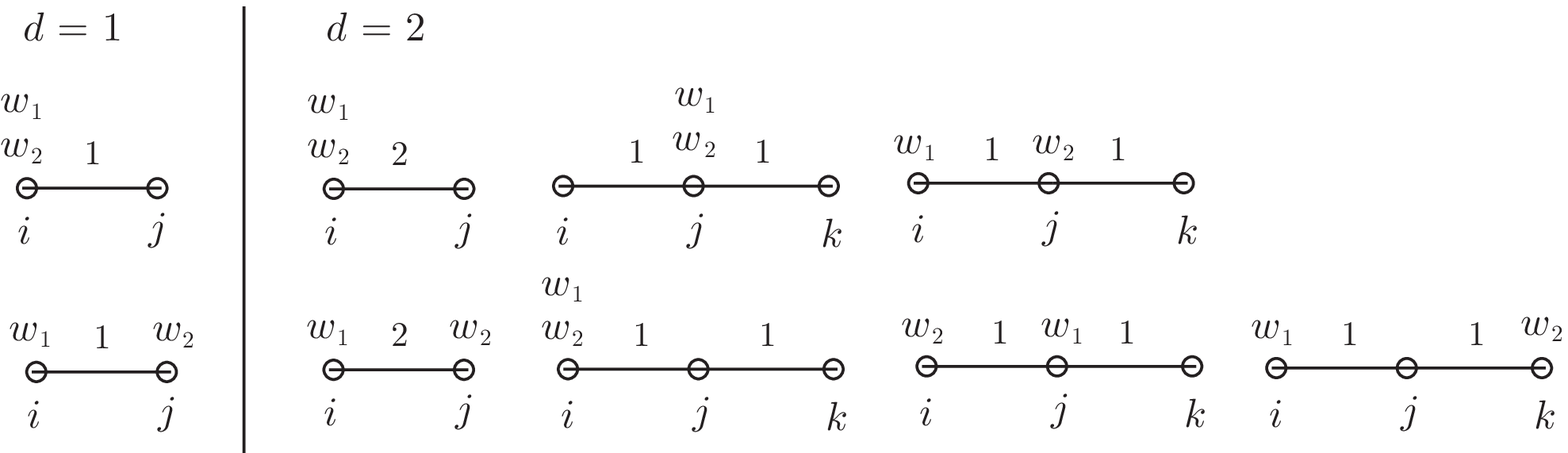}
 \caption{Tree graphs with two marked points $w_1, w_2$ up to degree $d=2$}
\label{closed_3}
\end{figure}

The Gromov-Witten invariants (\ref{closed_GWA}) can be computed by using the Atiyah-Bott fixed point theorem for the $T^{m+2}$-equivariant action on the embedding space ${\ICP}^{m+1}$. This 
torus action by $T^{m+2}=U(1)^{m+2}$ is equipped with the equivariant parameters $\lambda_1,\ldots,\lambda_{m+2}$. The $T^{m+2}$-equivariant action induces an action on $\overline{\cal M}_{0,n}({\ICP}^{m+1},d)$. The connected components of the fixed points of them are labeled by the tree graphs $\Gamma$ which consist of the vertices $v\in Vert(\Gamma)$ and the edges $e\in Edge(\Gamma)$ with degrees $d_e\in{\IN}$ whose summation is fixed by
\begin{equation}
d=\sum_{e\in Edge(\Gamma)}d_e.
\end{equation}
Here each vertex $v\in Vert(\Gamma)$ is labeled by $i_v\in \{1,\ldots,m+2\}$, and two different vertices $v,u$ connected by an edge $e\in Edge(\Gamma)$ should have different labels $i_v\neq  i_u$. In addition, each vertex $v\in Vert(\Gamma)$ is also colored by a set $S_v\subset \{1,\ldots,n\}$ induced by the marked points which satisfies
\begin{equation}
\{1,\ldots,n\}=\bigoplus_{v\in Vert(\Gamma)}S_v.
\label{marked_set}
\end{equation}
Some of the tree graphs with and without marked points are described in Figure \ref{closed_1}, \ref{closed_2}, and \ref{closed_3}.

Let $\overline{\cal M}_{\Gamma}$ be the set of fixed points in $\overline{\cal M}_{0,n}({\ICP}^{m+1},d)$ labeled by the tree graphs $\Gamma$, and  $\mathbf{e}\left(N_{\overline{\cal M}_{\Gamma}}\right)$ be the equivariant Euler class of the normal bundle to $\overline{\cal M}_{\Gamma}$. By the Atiyah-Bott fixed point theorem, one obtains \cite{Kontsevich:1994na}
\begin{align}
\begin{split}
&
\left<{\cal O}_{h^{p_1}}\cdots{\cal O}_{h^{p_n}}\right>_{0,d}^{{X}_{m+2}}=
\sum_{\Gamma}\frac{1}{\left|Aut(\Gamma)\right|}\int_{\overline{\cal M}_{\Gamma}}\frac{i^*_{\overline{\cal M}_{\Gamma}}\phi}{\mathbf{e}\left(N_{\overline{\cal M}_{\Gamma}}\right)},
\\
&\phi \equiv c_{(m+2)d+1}({\cal E}_d)ev^*_1(h^{p_1})\cdots ev^*_n(h^{p_n}),
\end{split}
\label{closed_GWA_lo}
\end{align}
where $i_{\overline{\cal M}_{\Gamma}}:\overline{\cal M}_{\Gamma}\hookrightarrow\overline{\cal M}_{0,n}({\ICP}^{m+1},d)$ is the inclusion, and
\begin{align}
\left|Aut(\Gamma)\right|&=\mbox{(order of the automorphism group of $\Gamma$)}\nonumber\\
&\ \ \ \
\times \mbox{(product of the degrees of edges)}.
\end{align}
The integration in (\ref{closed_GWA_lo}) is decomposed as
\begin{equation}
\int_{\overline{\cal M}_{\Gamma}}\frac{i^*_{\overline{\cal M}_{\Gamma}}\phi}{\mathbf{e}\left(N_{\overline{\cal M}_{\Gamma}}\right)}=
\prod_{e\in Edge(\Gamma)}E_{i,j}(\lambda_1,\ldots,\lambda_{m+2})\cdot \prod_{v\in Vert(\Gamma)}V_i(\lambda_1,\ldots,\lambda_{m+2}),
\label{each_closed_GWA_lo}
\end{equation}
where
\begin{equation}
E_{i,j}(\lambda_1,\ldots,\lambda_{m+2})=
\frac{\prod_{a=0}^{(m+2)d_e}\frac{a\lambda_i+\big((m+2)d_e-a\big)\lambda_j}{d_e}}{(-1)^{d_e}\frac{(d_e!)^2}{d_e^{2d_e}}(\lambda_i-\lambda_j)^{2d_e}\prod_{k\neq i,j}^{m+2}\prod_{a=0}^{d_e}\left(\frac{a\lambda_i+(d_e-a)\lambda_j}{d_e}-\lambda_k\right)}
\label{localize_Eij}
\end{equation}
is the contribution from an edge $e\in Edge(\Gamma)$ which connects two vertices labeled by $i,j$, and 
\begin{align}
V_i(\lambda_1,\ldots,\lambda_{m+2})&=
\frac{\lambda_i^{p_v}}{\big((m+2)\lambda_i\big)^{val(v)-1}}\prod_{j\neq i}^{m+2}(\lambda_i-\lambda_j)^{val(v)-1}\nonumber\\
&
\times \prod_{\mbox{\scriptsize flags}\ F=(v,\alpha)}\frac{d_{e_{\alpha}}}{\lambda_i-\lambda_{j_{\alpha}}}\cdot \left(\sum_{\mbox{\scriptsize flags}\ F=(v,\alpha)}\frac{d_{e_{\alpha}}}{\lambda_i-\lambda_{j_{\alpha}}}\right)^{val(v)-3+s_v}
\label{localize_Vi}
\end{align}
is the contribution from a vertex $v\in Vert(\Gamma)$ labeled by $i$. Here $val(v)$ is the valency of the vertex $v$ which means the number of flags $F=(v,\alpha)$ associated with the vertex $v$. Each flag $F=(v,\alpha)$ consists of an edge $e_{\alpha}$ which connects two vertices $v$ and $\alpha$ labeled by $j_{\alpha}$. The variable $s_v$ is the cardinality of $S_v=\{w_1,\ldots,w_{s_v}\}\subset\{1,\ldots,n\}$, where $\sum_{v\in Vert(\Gamma)}s_v=n$ from (\ref{marked_set}), and
\begin{equation}
p_v=\sum_{a=1}^{s_v}p_{w_a}
\end{equation}
for $\left<{\cal O}_{h^{p_1}}\cdots{\cal O}_{h^{p_n}}\right>_{0,d}^{{X}_{m+2}}$. By (\ref{closed_GWA_lo}), one can compute the total number of degree $d$ rational curves in the Calabi-Yau $m$-fold $X_{m+2}$. Even though the expression (\ref{each_closed_GWA_lo}) depends on the equivariant parameters, by summing up all possible tree graphs in (\ref{closed_GWA_lo}), one obtains rational numbers.

Let us consider the generating function of the Gromov-Witten invariants (\ref{closed_GWA}) given by the $n$-point function on ${\ICP}^1$:
\begin{equation}
\left<{\cal O}_{h^{p_1}}\cdots{\cal O}_{h^{p_n}}\right>_{{\ICP}^1}^{{X}_{m+2}}
\Big|_{\scriptsize\mbox{instantons}}\equiv
\sum_{d=1}^{\infty}\left<{\cal O}_{h^{p_1}}\cdots{\cal O}_{h^{p_n}}\right>_{0,d}^{{X}_{m+2}}q^d.
\end{equation}
It was conjectured in \cite{Klemm:2007in} that the multiple covering formula for this kind of $n$-point function is given by
\begin{equation}
\left<{\cal O}_{h^{p_1}}\cdots{\cal O}_{h^{p_n}}\right>_{{\ICP}^1}^{{X}_{m+2}}
\Big|_{\scriptsize\mbox{instantons}}=
\sum_{d=1}^{\infty}n_d(h^{p_1},\ldots,h^{p_n})\sum_{k=1}^{\infty}k^{-3+n}q^{kd},
\end{equation}
where $n_d(h^{p_1},\ldots,h^{p_n})$ take integer values. The results for some one- and two-point functions for $m$-dimensional Calabi-Yau manifolds are listed in Table \ref{GW_cy7_x9} (for $m=7$), Table \ref{GW_cy9_x11} (for $m=9$), and 
Table \ref{GW_cy11_x13} (for $m=11$).
\begin{table}[t]
\begin{center}
\begin{tabular}{|c|r|r|r|}
\hline
$d$ & $n_{d}(h^{5})$ & $n_{d}(h^{2},h^{4})$ & $n_{d}(h^{3},h^{3})$ \\ \hline
1 & 253490796 & 763954092 & 1069047153 \\
2 & 9757818340659360 & 93777295128674544 & 156037426159482684 \\
3 & 897560654227562339370036 & 17873898563070361396216980 & 33815935806268253433549768 \\ \hline
\end{tabular}
\caption{Closed Gromov-Witten invariants for the Calabi-Yau sevenfold $X_9$}
\label{GW_cy7_x9}
\end{center}
\end{table}
\begin{table}[t]
\begin{center}
\begin{tabular}{|c|r|r|}
\hline
$d$ & $n_{d}(h^{7})$ & $n_{d}(h^{2},h^{6})$ \\ \hline
1 & 69407571816 & 307393401172 \\
2 & 1141331429965005402156 & 18502500911901094328812 \\
3 & 54092342700646414182478412213976 & 1956023268644595599036553129234708 \\ \hline
$d$ & $n_{d}(h^{3},h^{5})$ & $n_{d}(h^{4},h^{4})$ \\ \hline
1 & 695221679878 & 905702054829 \\
2 & 63961167525267587307458 & 96846834660195439038593 \\
3 & 9166866145651267306745680150978774 & 15543750432092710582980268877325624 \\ \hline
\end{tabular}
\caption{Closed Gromov-Witten invariants for the Calabi-Yau ninefold $X_{11}$}
\label{GW_cy9_x11}
\end{center}
\end{table}
\begin{table}[t]
\begin{center}
\begin{tabular}{|c|r|}
\hline
$d$ & $n_{d}(h^{9})$ \\ \hline
1 & 22292891367552 \\
2 & 213784310853904983595932672 \\
3 & 7334754565028499250611499436120804633472 \\ \hline
$d$ & $n_{d}(h^{2},h^{8})$ \\ \hline
1 & 132851852317704 \\
2 & 5016156939319858820675234112 \\
3 & 393681022044306633159040293084738738617256 \\ \hline
$d$ & $n_{d}(h^{3},h^{7})$ \\ \hline
1 & 428548261855948 \\
2 & 28864547587341497987819790784 \\
3 & 3354492048909382182232089138021915233392428 \\ \hline
$d$ & $n_{d}(h^{4},h^{6})$ \\ \hline
1 & 844604779452382 \\
2 & 83333583532522526645403196872 \\
3 & 12807226634327588945692825517163202349337982 \\ \hline
$d$ & $n_{d}(h^{5},h^{5})$ \\ \hline
1 & 1055381615783157 \\
2 & 118907743429844203896833928442 \\
3 & 20218987604455477572463422690122440316972424 \\ \hline
\end{tabular}
\caption{Closed Gromov-Witten invariants for the Calabi-Yau elevenfold $X_{13}$}
\label{GW_cy11_x13}
\end{center}
\end{table}

\subsection{Counting holomorphic disks}\label{subapp:localization_open}

In the following, we consider the odd dimensional Calabi-Yau hypersurface $X_{m+2}$ with
\begin{equation}
m=2\mathfrak{m}+1.
\end{equation}
Let $L_{m+2}=X_{m+2}^{\IR}$ be a Lagrangian submanifold in $X_{m+2}$ which is the real locus defined by the fixed points of an anti-holomorphic involution on ${\ICP}^{m+1}$:
\begin{equation}
x_{2i-1}\ \mapsto\ \overline{x}_{2i},\ \ \
x_{2i}\ \mapsto\ \overline{x}_{2i-1},\ \ \
x_{2\mathfrak{m}+3}\ \mapsto\ \overline{x}_{2\mathfrak{m}+3},
\label{real_anti_fix}
\end{equation}
where $i=1,\ldots,\mathfrak{m}+1$. The degree $d$ disk Gromov-Witten invariant (see \cite{Iacovino:2009xf, Iacovino:2009aj} for a mathematical detail) of $X_{m+2}$ is defined by
\begin{equation}
\left<{\cal O}_{h^{p_1}}\cdots{\cal O}_{h^{p_n}}\right>_{D,d}^{{X}_{m+2}} \equiv
\int_{\widetilde{\cal M}_{D,n}({\ICP}^{m+1}/{\ICP}^{m+1}_{\IR},d)}\mathbf{e}({\cal E}_d^{\IR})ev^*_1(h^{p_1})\cdots ev^*_n(h^{p_n}).
\label{open_GWA}
\end{equation}
Here $\widetilde{\cal M}_{D,n}({\ICP}^{m+1}/{\ICP}^{m+1}_{\IR},d)$ is the moduli space of degree $d$ and $n$-pointed stable disk maps to ${\ICP}^{m+1}$, where the boundary of the disk $\partial D$ is mapped to the Lagrangian ${\ICP}^{m+1}_{\IR}$ obtained as the set of fixed points of (\ref{real_anti_fix}). $\mathbf{e}({\cal E}_d^{\IR})$ is the Euler class of a real vector bundle ${\cal E}_d^{\IR}$ on $\widetilde{\cal M}_{D,n}({\ICP}^{m+1}/{\ICP}^{m+1}_{\IR},d)$ with real rank $(m+2)d+1$. We do not give a mathematically rigorous definition of (\ref{open_GWA}). Instead, as in \cite{Jinzenji:2011vm}, we follow the definition in \cite{Pandharipande:2006d} for the quintic Calabi-Yau threefold, and assume that their analysis can be straightforwardly extended into higher dimensions. As in (\ref{select_rule}), then one finds the following condition for obtaining the nontrivial quantity:
\begin{equation}
\sum_{i=1}^np_i=\frac{m-3}{2}+n.
\label{op_select_rule}
\end{equation}

\begin{figure}[t]
 \centering
  \includegraphics[width=125mm]{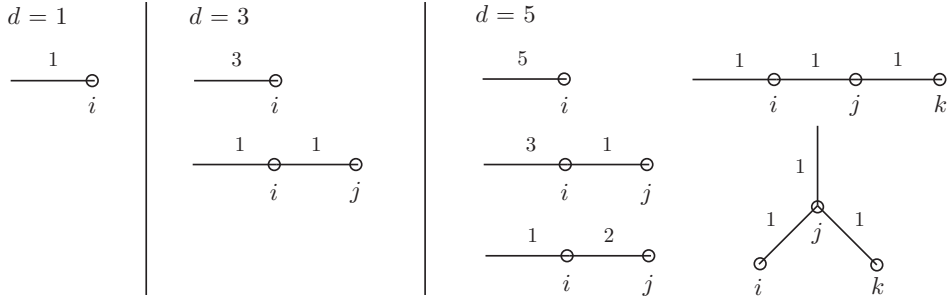}
 \caption{Open tree graphs without marked points up to degree $d=5$}
\label{open_1}
\end{figure}

\begin{figure}[t]
 \centering
  \includegraphics[width=125mm]{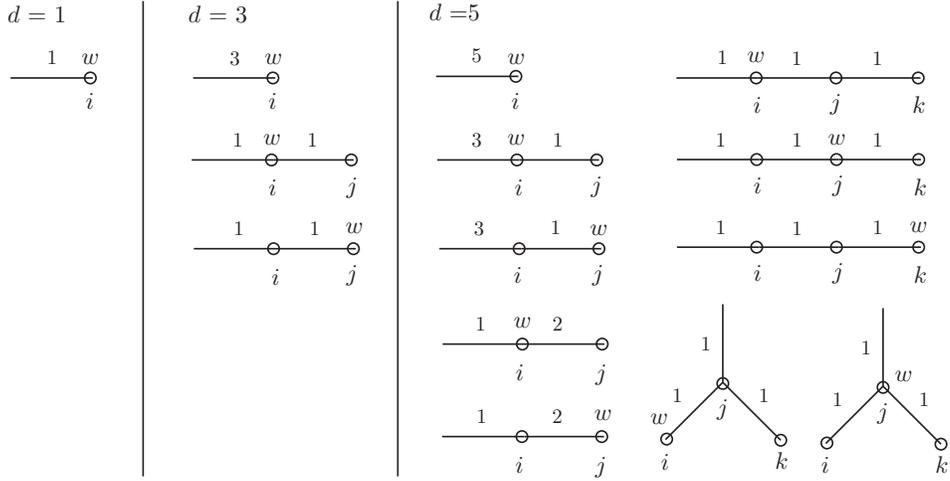}
 \caption{Open tree graphs with one marked point $w$ up to degree $d=5$}
\label{open_2}
\end{figure}

\begin{figure}[t]
 \centering
  \includegraphics[width=165mm]{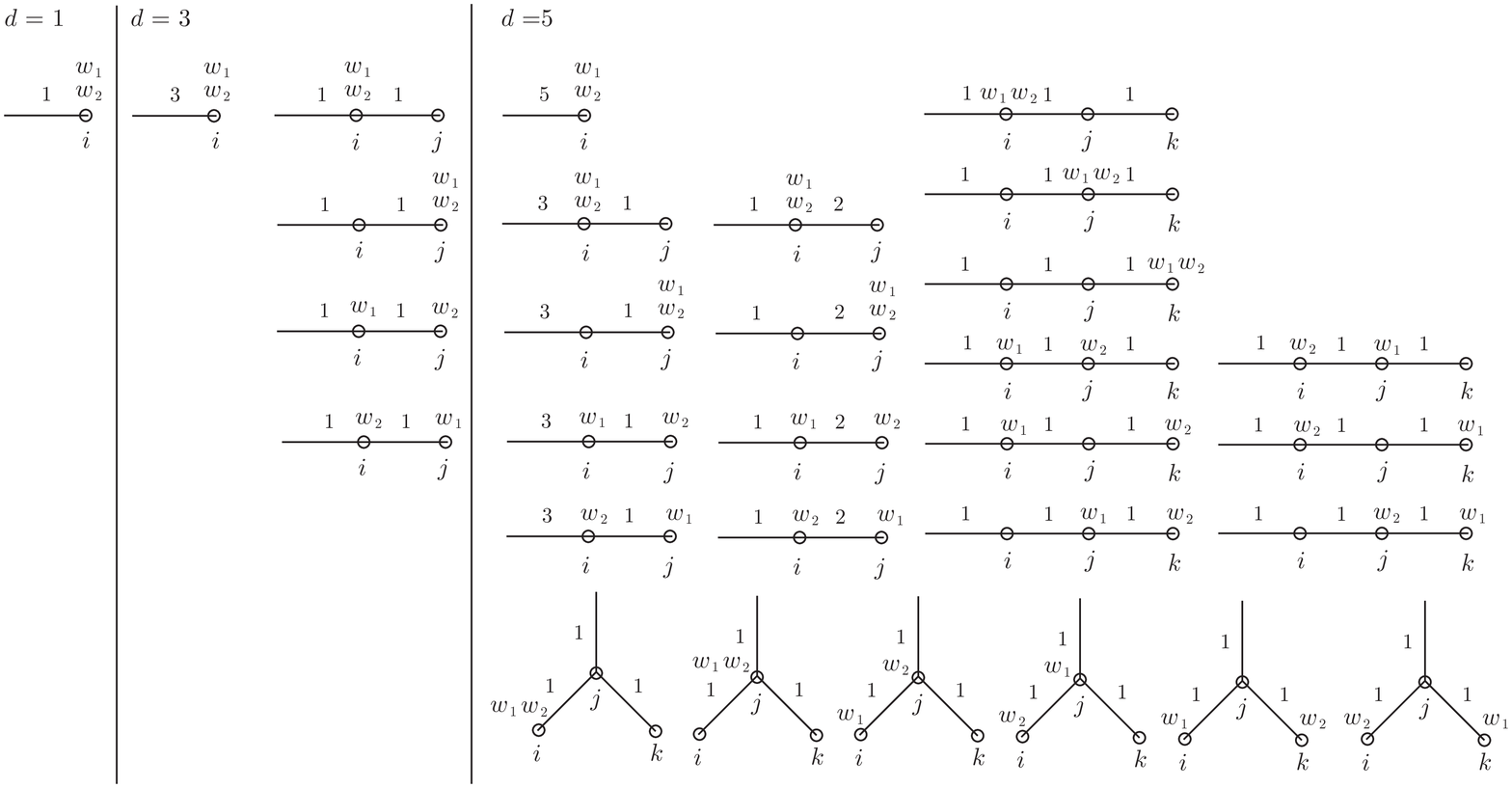}
 \caption{Open tree graphs with two marked points $w_1, w_2$ up to degree $d=5$}
\label{open_3}
\end{figure}

As was described in Appendix \ref{subapp:localization_close}, let us apply the Atiyah-Bott fixed point theorem to compute the disk Gromov-Witten invariant (\ref{open_GWA}). As performed in \cite{Pandharipande:2006d}, we consider a subtorus $\widetilde{T}^{\mathfrak{m}+1}\subset T^{m+2}$ which preserves the Lagrangian $L_{m+2}$ as
\begin{equation}
\widetilde{T}^{\mathfrak{m}+1}:\ \ 
x_{2i-1}\ \mapsto\ \xi_i x_{2i-1},\ \ \
x_{2i}\ \mapsto\ \overline{\xi}_i x_{2i}, \ \ \
x_{2\mathfrak{m}+3}\ \mapsto\ x_{2\mathfrak{m}+3},
\end{equation}
where $i=1,\ldots,\mathfrak{m}+1$. The $\widetilde{T}^{\mathfrak{m}+1}$-equivariant parameters $\widetilde{\lambda}_1,\ldots,\widetilde{\lambda}_{\mathfrak{m}+1}$ on $L_{m+2}$ are defined by \cite{Pandharipande:2006d}
\begin{equation}
\rho^*(\lambda_{2i-1})=-\rho^*(\lambda_{2i})=\widetilde{\lambda}_i,\ \ \ \
\rho^*(\lambda_{m+2})=0,
\label{restrict_equiv}
\end{equation}
where
\begin{equation}
\rho:\ {\IQ}[\widetilde{\lambda}_1,\ldots,\widetilde{\lambda}_{\mathfrak{m}+1}]\
\hookrightarrow\ 
{\IQ}[\lambda_1,\ldots,\lambda_{m+2}].
\end{equation}
The $\widetilde{T}^{\mathfrak{m}+1}$-equivariant action induces an action on 
$\widetilde{\cal M}_{D,n}({\ICP}^{m+1}/{\ICP}^{m+1}_{\IR},d)$, and the connected components of the fixed points are labeled by open tree graphs $\widetilde{\Gamma}$. Each tree graph $\widetilde{\Gamma}$ consists of vertices $v\in Vert(\widetilde{\Gamma})$, edges $e\in Edge(\widetilde{\Gamma})$ with degrees $d_e\in{\IN}$, and also a half-edge $\widehat{e}$ with an odd degree $d_0$. The total degree of $\widetilde{\Gamma}$ is given by
\begin{equation}
d=d_0+2\sum_{e\in Edge(\widetilde{\Gamma})}d_e.
\end{equation}
The each vertex $v\in Vert(\widetilde{\Gamma})$ is labeled by $i_v\in \{1,\ldots,m+2\}$ and also colored by a set $S_v\subset \{1,\ldots,n\}$ in (\ref{marked_set}). The each edge $e\in Edge(\widetilde{\Gamma})$ connects two different vertices $v,u$ with different labels $i_v\neq  i_u$, whereas the half-edge $\widehat{e}$ have only one vertex $v$. Some of the open tree graphs with and without marked points are described in Figure \ref{open_1}, \ref{open_2}, and \ref{open_3}.

Then as discussed in \cite{Jinzenji:2011vm, Popa:2013jsa} (see also \cite{Jinzenji:2013uqa}), one expects
\begin{align}
\begin{split}
&
\left<{\cal O}_{h^{p_1}}\cdots{\cal O}_{h^{p_n}}\right>_{D,d}^{{X}_{m+2}}=
\sum_{\widehat{\Gamma}}\frac{2^n}{\big|Aut(\widetilde{\Gamma})\big|}\int_{\overline{\cal M}_{\widetilde{\Gamma}}}\frac{\mathfrak{i}^*_{\overline{\cal M}_{\widetilde{\Gamma}}}\varphi}{\mathbf{e}\big(N_{\overline{\cal M}_{\widetilde{\Gamma}}}\big)},
\\
&
\varphi\equiv \mathbf{e}({\cal E}_d^{\IR})ev^*_1(h^{p_1})\cdots ev^*_n(h^{p_n}),
\end{split}
\label{open_GWA_lo}
\end{align}
where $\mathfrak{i}_{\overline{\cal M}_{\widetilde{\Gamma}}}:\overline{\cal M}_{\widetilde{\Gamma}}\hookrightarrow\widetilde{\cal M}_{D,n}({\ICP}^{m+1}/{\ICP}^{m+1}_{\IR},d)$ is the inclusion.
In analogy with \cite{Walcher:2006rs, Pandharipande:2006d}, the integration in (\ref{open_GWA_lo}) is supposed to be decomposed as
\begin{align}
\int_{\overline{\cal M}_{\widetilde{\Gamma}}}\frac{\mathfrak{i}^*_{\overline{\cal M}_{\widetilde{\Gamma}}}\varphi}{\mathbf{e}\big(N_{\overline{\cal M}_{\widetilde{\Gamma}}}\big)}&=
\prod_{e\in Edge(\widetilde{\Gamma})}E_{i,j}(\lambda_1,\ldots,\lambda_{m+2})\cdot \prod_{v\in Vert(\widetilde{\Gamma})}V_i(\lambda_1,\ldots,\lambda_{m+2})
\nonumber\\
&\ \ \times
F_{\ell}(\lambda_1,\ldots,\lambda_{2\mathfrak{m}+2}),
\label{each_open_GWA_lo}
\end{align}
where $E_{i,j}(\lambda_1,\ldots,\lambda_{m+2})$ and $V_i(\lambda_1,\ldots,\lambda_{m+2})$ are defined in (\ref{localize_Eij}) and (\ref{localize_Vi}), respectively, and
\begin{equation}
F_{\ell}(\lambda_1,\ldots,\lambda_{2\mathfrak{m}+2})=
\frac{(-1)^{\mathfrak{m}-1}\prod_{a=0}^{\frac{(m+2)d_0-1}{2}}\frac{a\lambda_{\ell}+\big((m+2)d_0-a\big)\lambda_{\sigma(\ell)}}{d_0}}{(-1)^{\frac{d_0-1}{2}}\frac{d_0!}{d_0^{d_0}}(\lambda_{\ell}-\lambda_{\sigma(\ell)})^{d_0}\prod_{k\neq \ell,\sigma(\ell)}^{2\mathfrak{m}+2}\prod_{a=0}^{\frac{d_0-1}{2}}\left(\frac{a\lambda_{\ell}+(d_0-a)\lambda_{\sigma(\ell)}}{d_0}-\lambda_k\right)},
\label{localize_Fell}
\end{equation}
for the half-edge $\widehat{e}$ with a vertex labeled by $\ell$ in $\widetilde{\Gamma}$. Taking into account the relation in (\ref{restrict_equiv}), the equivariant parameters in (\ref{each_open_GWA_lo}) should be restricted as 
\begin{equation}
\lambda_{2i-1}=-\lambda_{2i},\ \ \ \
\lambda_{m+2}=0.
\end{equation}
Accordingly, for the vertex labeled by $\ell$ with the half-edge $\widehat{e}$ in (\ref{localize_Vi}), we need to consider that $\widehat{e}$ in the flag connects the vertex labelled by $\ell$ and the virtual vertex labelled by $\sigma(\ell)$, where $\sigma(2i-1)=2i$, $\sigma(2i)=2i-1$, and $\sigma(m+2)=m+2$.

The $n$-point function on the disk $D$ gives the generating function of the open Gromov-Witten invariants (\ref{open_GWA}) as
\begin{equation}
\left<{\cal O}_{h^{p_1}}\cdots{\cal O}_{h^{p_n}}\right>_{D}^{{X}_{m+2}}
\Big|_{\scriptsize\mbox{instantons}}\equiv
\sum_{d=1}^{\infty}\left<{\cal O}_{h^{p_1}}\cdots{\cal O}_{h^{p_n}}\right>_{D,2d-1}^{{X}_{m+2}}q^{\frac{2d-1}{2}}.
\end{equation}
For the multiple covering formula, it is natural to consider that one can generalize the formula for the one-point function in (\ref{multi_cov_cym}) into the form
\begin{align}
&
\left<{\cal O}_{h^{p_1}}\cdots{\cal O}_{h^{p_n}}\right>_{D}^{{X}_{m+2}}
\Big|_{\scriptsize\mbox{instantons}}&
\nonumber\\
&=
\sum_{d=1}^{\infty}n_{2d-1}^{open}(h^{p_1},\ldots,h^{p_n})\sum_{k=1}^{\infty}(-1)^{(\mathfrak{m}-1)(k-1)}(2k-1)^{-2+n}q^{\frac{(2d-1)(2k-1)}{2}},
\end{align}
where $n_{2d-1}^{open}(h^{p_1},\ldots,h^{p_n})$ take integer values. The results for some one- and two-point functions for $m$-dimensional Calabi-Yau manifold are listed in Table \ref{o_GW_cy7_x9} (for $m=7$), Table \ref{o_GW_cy9_x11} (for $m=9$), and Table \ref{o_GW_cy11_x13} (for $m=11$).
\begin{table}[t]
\begin{center}
\begin{tabular}{|c|r|r|}
\hline
$d$ & $n_{d}^{open}(h^{3})$ & $n_{d}^{open}(h^{2},h^{2})$ \\ \hline
1 & 1890 & 1890 \\
3 & 94563624960 & 390888167040 \\
5 & 16211885196706741080 & 125320461870005007480 \\ \hline
\end{tabular}
\caption{Open Gromov-Witten invariants for the Calabi-Yau sevenfold $X_9$}
\label{o_GW_cy7_x9}
\end{center}
\end{table}
\begin{table}[t]
\begin{center}
\begin{tabular}{|c|r|r|}
\hline
$d$ & $n_{d}^{open}(h^{4})$ & $n_{d}^{open}(h^{2},h^{3})$ \\ \hline
1 & 20790 & 20790 \\
3 & 739689094281060 & 3401855947080360 \\
5 & 92349241505201808072653400 & 813675420905155787309922360 \\ \hline
\end{tabular}
\caption{Open Gromov-Witten invariants for the Calabi-Yau ninefold $X_{11}$}
\label{o_GW_cy9_x11}
\end{center}
\end{table}
\begin{table}[t]
\begin{center}
\begin{tabular}{|c|r|}
\hline
$d$ & $n_{d}^{open}(h^{5})$ \\ \hline
1 & 270270 \\
3 & 9630787776863673420 \\
5 & 1259056659533544456991412149863720 \\
\hline
$d$ & $n_{d}^{open}(h^{2},h^{4})$ \\ \hline
1 & 270270 \\
3 & 45988227483874635960 \\
5 & 11540778289906173258398255936168040 \\
\hline
$d$ & $n_{d}^{open}(h^{3},h^{3})$ \\ \hline
1 & 270270 \\
3 & 52135755838174728720 \\
5 & 13515613082202105997576568972577720 \\ \hline
\end{tabular}
\caption{Open Gromov-Witten invariants for the Calabi-Yau elevenfold $X_{13}$}
\label{o_GW_cy11_x13}
\end{center}
\end{table}

\section{Monodromy analysis}\label{app:monodromy}

Following the prescription of \cite{Walcher:2006rs}, we will consider the analytic continuation of the inhomogeneous solution (\ref{s_pot_gen_m}) for the septic Calabi-Yau fivefold which takes the form
\begin{equation}
c^{-1} \tau (z)=
\frac{\pi^3}{4} \sum_{n=0}^{\infty}\frac{\Gamma \left(7n+\frac{9}{2}\right)}
{\Gamma \left(n+\frac{3}{2}\right)^7}z^{n+\frac{1}{2}}.
\label{7F6}
\end{equation}
To perform the analytic continuation, it is convenient to use the Barnes type integral representation as 
\begin{align}
c^{-1} \tau (z)=\frac{\pi^3}{4} \frac{1}{2 \pi i}\int_{-i\infty}^{i\infty} \frac{\Gamma{\left(-s+\frac{1}{2}\right)}\Gamma{(7s+1)}\Gamma{(s+\frac{1}{2}})}{\Gamma{(s+1)}^7} e^{i \pi (s-\frac{1}{2})}z^s ds.
\end{align}
By closing the contour on the right half plane, we can reproduce the expression in (\ref{7F6}). Meanwhile, if we close the contour on the left half plane, we obtain
\begin{align}
c^{-1} \tau (z)=\frac{\pi^3}{4}\Bigg[ \sum_{n=0}^{\infty} \frac{-\Gamma{(-7n-\frac{5}{2})}}{\Gamma{(-n+\frac{1}{2})^7}}z^{-n-\frac{1}{2}}+\sum_{n=1}^{\infty} \frac{-\Gamma{(\frac{n}{7})}e^{6 \pi i n/7}}{7\Gamma{(n)}\Gamma{(1-\frac{n}{7})^6}}z^{-\frac{n}{7}}  
e^{-\frac{i\pi}{2}}\frac{\sin{\frac{\pi n}{7}}}{\cos{\frac{\pi n}{7}}} \Bigg].
\label{LHPtau}
\end{align}
Note that the residue of $\Gamma{(ks+1)}$ at pole $-\frac{n}{k}$ is $\frac{(-1)^{n-1}}{k \Gamma{(n)}}$. 
The first term in (\ref{LHPtau}) changes the overall sign under the monodromy transformation $z^{\frac{1}{7}} \rightarrow e^{-\frac{2\pi i}{7}} z^{\frac{1}{7}}$. On the other hand, the second term in (\ref{LHPtau}) is a solution of the ordinary Picard-Fuchs equation for the closed string sector (\ref{closed_PF}) around the Landau-Ginzburg point $z=\infty$.

By using the basis of the closed string periods around $z=\infty$ given by
\begin{align}
\varpi_j (z)=\sum_{n=1}^{\infty}\frac{-\Gamma{(\frac{n}{7})}e^{6 \pi i n/7}}{7 \Gamma{(n)}\Gamma{\left(1-\frac{n}{7}\right)^6}}z^{-\frac{n}{7}}e^{2 \pi i j n/7}, \ \ \ \ \ \ \ j=0, \ldots, 6
\end{align}
satisfying $\sum_j \varpi_j (z)=0$ and the identity
\begin{align}
\frac{\sin{\frac{\pi n}{7}}}{\cos{\frac{\pi n}{7}}}=2\sin{\frac{2\pi n}{7}}-2\sin{\frac{4\pi n}{7}}+2\sin{\frac{6\pi n}{7}},
\end{align}
we find that the second term in (\ref{LHPtau}) can be represented by
\begin{align}
\frac{\pi^3}{4}(\varpi_0+2 \varpi_2+2\varpi_4+2\varpi_6 ).
\label{LGtau2}
\end{align}
In the case of septic fivefold, the Landau-Ginzburg basis $\varpi =(\varpi_0, \varpi_1, \varpi_2, \varpi_3, \varpi_5, \varpi_6)^T$ is related to the large volume basis $\Pi=(\Pi_0, \Pi_1, \Pi_2, \Pi_3, \Pi_4, \Pi_5)^T$ by (see, for example \cite{Sugiyama:2000aw})
\begin{align}
\Pi=\begin{pmatrix}
1& 0 & 0 & 0 & 0 & 0 \\
-\frac{3}{7} & \frac{3}{7} & \frac{2}{7} & \frac{1}{7} & -\frac{1}{7} & -\frac{2}{7} \\
\frac{3}{7} & \frac{6}{7} & \frac{3}{7} & \frac{1}{7} & 0 & \frac{1}{7} \\
-4 & 4 & 1 & 0 & 0 & -1 \\
-8 & -4 & -3 & -1 & -1 & -4 \\
1 & -1 & 0 & 0 & 0 & 0
\end{pmatrix}\varpi ,
\end{align}
thus we can also express (\ref{LGtau2}) as
\begin{align}
\frac{\pi^3}{4}(63\Pi_0+98 \Pi_1-196 \Pi_2+8\Pi_3-16\Pi_4-32\Pi_5 ).
\end{align}
In terms of the large volume basis, the monodromy matrix around the Landau-Ginzburg point is given by \cite{Sugiyama:2000aw}
\begin{align}
M=\begin{pmatrix}
1& 0 & 0 & 0 & 0 & -1 \\
-1& 1 & 0 & 0 & 0 & 1 \\
-1& 1 & 1 & 0 & 0 & 1 \\
-14 & 0 & 7 & 1 & 0 & 14 \\
-7 & -14 & 7 & 1 & 1 & 7 \\
7 & 7 & -14 & 0 & -1 & -6
\end{pmatrix}.
\end{align}
As a result, the monodromy transformations of the solution $\tau (z)$ under $z^{\frac{1}{7}} \rightarrow e^{-\frac{2\pi i}{7}} z^{\frac{1}{7}}$ is given by
\begin{align}
\tau \rightarrow -\tau-\frac{\pi^3}{4}\Pi_5.
\end{align}

Suppose that the normal function associated with $C_{+}-C_{-}$ (see \cite{Morrison:2007bm} for details) is defined by
\begin{align}
{\cal{T}}_{\pm}(z)=\frac{\Pi_1}{2}\pm\frac{\Pi_0}{4}\pm c^{-1} \tau (z).
\end{align}
Since the Landau-Ginzburg monodromy transforms the classical terms as $\frac{\Pi_1}{2}+\frac{\Pi_0}{4} \rightarrow 
\frac{\Pi_1}{2}-\frac{\Pi_0}{4}+\frac{\Pi_5}{4}$, we find that the natural choice for the normalization factor is $c=\pi^3$.
This result is consistent with the observation in the Appendix of \cite{Jinzenji:2011vm}.

\section{On-shell disk two-point function}\label{app:disk_two}

In this appendix, we will verify the formula (\ref{two_pt_disk_formula}) for the $m=2\mathfrak{m}+1$ dimensional Calabi-Yau hypersurface $X_{m+2}$:
\begin{equation}
\big<{\cal O}_{h^p}{\cal O}_{h^{\mathfrak{m}-p+1}}\big>_{D}^{X_{m+2}}=
\prod_{k=1}^{p-1}
\frac{
\big<{\cal O}_{h}{\cal O}_{h^{\mathfrak{m}-p+k}}{\cal O}_{h^{\mathfrak{m}+p-k}}\big>_{\mathrm{\ICP}^1}^{X_{m+2}}}
{\big<{\cal O}_{h}{\cal O}_{h^{k}}{\cal O}_{h^{2\mathfrak{m}-k}}\big>_{\mathrm{\ICP}^1}^{X_{m+2}}}
\times
\big<{\cal O}_h{\cal O}_{h^{\mathfrak{m}}}\big>_{D}^{X_{m+2}}.
\label{two_pt_disk_formula_ap}
\end{equation}
Note that to obtain the expression (\ref{two_pt_disk_formula}), we need to use the divisor equation.

Let $\widetilde{\cal O}_{p}$ be the observable in the mirror B-model on $Y_{m+2}$ corresponding to the observable ${\cal O}_{h^p}$ in the A-model on $X_{m+2}$. These operators obey the fusion rule \cite{Greene:1993vm, Klemm:1996ts}
\begin{equation}
\widetilde{\cal O}_{p}\widetilde{\cal O}_{q}=\eta \widetilde{Y}_q^p
\widetilde{\cal O}_{p+q},\ \ \ \
\widetilde{Y}_q^p\equiv \big<\widetilde{\cal O}_p\widetilde{\cal O}_q\widetilde{\cal O}_{m-p-q}\big>_{\mathrm{\ICP}^1}^{Y_{m+2}},
\end{equation}
where $\eta$ is a constant. Two different factorizations of a four-point function by the fusion rule yield a relation \cite{Greene:1993vm, Klemm:1996ts}
\begin{equation}
\widetilde{Y}_q^p=\frac{\prod_{k=0}^{p-1}\widetilde{Y}_{q+k}^1}{\prod_{k=1}^{p-1}\widetilde{Y}_k^1}.
\end{equation}
From the fusion rule we have
\begin{equation}
\widetilde{\cal O}_{p}\widetilde{\cal O}_{q}=\eta \widetilde{Y}_{q}^p\widetilde{\cal O}_{p+q}
=\frac{\prod_{k=0}^{p-2}\widetilde{Y}_{q+k}^1}{\prod_{k=1}^{p-1}\widetilde{Y}_k^1}
\times
\eta\widetilde{Y}_{p+q-1}^1\widetilde{\cal O}_{p+q}
=\prod_{k=1}^{p-1}
\frac{\widetilde{Y}_{q+k-1}^1}{\widetilde{Y}_k^1}
\times
\widetilde{\cal O}_{1}\widetilde{\cal O}_{p+q-1}.
\end{equation}
Then we can expect a relation for disk two-point functions in the B-model
\begin{equation}
\big<\widetilde{\cal O}_{p}\widetilde{\cal O}_{q}\big>_{D}^{Y_{m+2}}=
\prod_{k=1}^{p-1}
\frac{\widetilde{Y}_{q+k-1}^1}{\widetilde{Y}_k^1}
\times
\big<\widetilde{\cal O}_{1}\widetilde{\cal O}_{p+q-1}\big>_{D}^{Y_{m+2}}.
\end{equation}
For $q=\mathfrak{m}-p+1$, this relation is nothing but the mirror dual of (\ref{two_pt_disk_formula_ap}).

\section{Solutions to the extended GKZ system}\label{app:sol_ext_gkz}

For the mirror Calabi-Yau hypersurfaces $Y_{m+2}$, the solutions of the extended GKZ operators $\{\widetilde{\mathcal{D}}_1, \mathcal{D}_2, \widetilde{\mathcal{D}}_{3}\}$ constructed in Section \ref{subsec:disk_glsm_ext_GKZ} can be obtained by the Frobenius method. 

Starting with a function
\begin{align}
\begin{split}
\Pi(z_1,z_2;\rho_1,\rho_2)&
=\sum_{k_1,k_2=0}^{\infty}
\frac{\Gamma\big((m+1)(k_1+\rho_1)+(k_2+\rho_2)+1\big)}
{\Gamma(k_1+\rho_1+1)^{m+1}\Gamma(k_2+\rho_2+1)} \\
&\times
\frac{z_1^{k_1+\rho_1}z_2^{k_2+\rho_2}}{\Gamma\big(-(k_1+\rho_1)+(k_2+\rho_2)+1\big)\Gamma\big((k_1+\rho_1)-(k_2+\rho_2)+1\big)},
\end{split}
\end{align}
we see that $2m+1$ independent solutions to $\widetilde{\mathcal{D}}_1\Pi=
\mathcal{D}_2\Pi=\widetilde{\mathcal{D}}_{3}\Pi=0$ corresponding to the variation of mixed Hodge structure (\ref{mixedH}) are given by
\begin{equation}
\Pi(z_1,z_2;0,0),\ \ \ \
\frac{\partial^p}{\partial \rho_1^p}\left.\Pi(z_1,z_2;\rho_1,0)\right|_{\rho_1=0},\ \ \ \
\Big(\frac{\partial}{\partial \rho_1}+\frac{\partial}{\partial \rho_2}\Big)^p
\left.\Pi(z_1,z_2;\rho_1,\rho_2)\right|_{\rho_1=\rho_2=0},
\end{equation}
where $p=1,\ldots,m$. The classical K\"ahler moduli $t_1$ and $t_2$ of the A-model are obtained by logarithmic solutions as
\begin{equation}
2\pi i t_1=\frac{1}{\Pi_0(z_1,z_2)}\frac{\partial}{\partial \rho_1}\left.\Pi(z_1,z_2;\rho_1,0)\right|_{\rho_1=0},\ \ \
2\pi i t_2=\frac{1}{\Pi_0(z_1,z_2)}\frac{\partial}{\partial \rho_2}\left.\Pi(z_1,z_2;0,\rho_2)\right|_{\rho_2=0},
\label{mirror_m_Xm}
\end{equation}
where  we defined a fundamental period as $\Pi_0(z_1,z_2) \equiv \Pi(z_1,z_2;0,0)$.
\begin{table}[t]
\begin{center}
\begin{tabular}{|c|rrrrr|}
\hline
$n_{d_1,d_2,1}$ & $d_1=0$ & 1 & 2 & 3 & 4 \\ \hline
$d_2=0$ &  & -33264 & 179259696 & -2237571970704 & 40027599733373280 \\
1 & 42 & 656208 & -3581843328 & 55738918811520 & -1223185862761241400 \\
2 & 0 & 542808 & 79461962496 & -856482988616640 & 20537586624138059952 \\
3 & 0 & -546588 & 81253624284 & 19846726854555888 & -284642465082679915848 \\
4 & 0 & 378168 & -139738847220 & 21927045021430752 & 6747123319002262489104 \\ 
\hline
$n_{d_1,d_2,2}$ & $d_1=0$ & 1 & 2 & 3 & 4 \\ \hline
$d_2=0$ &  & 33264 & -179259696 & 2237571970704 & -40027599733373280 \\
1 & -42 & 353584 & 3581843328 & -55738918811520 & 1223185862761241400 \\
2 & 0 & -542808 & 42777823592 & 856482988616640 & -20537586624138059952 \\
3 & 0 & 546588 & -81253624284 & 10681944890924216 & 284642465082679915848 \\
4 & 0 & -378168 & 139738847220 & -21927045021430752 & 3631071746533555621612 \\ 
\hline
\end{tabular}
\caption{Off-shell open Gromov-Witten invariants for the septic Calabi-Yau fivefold. Note that $n_{d_1,d_2,1}=-n_{d_1,d_2,2}$ for $d_1\neq d_2$ and a linear combination $n_{d,d,1}+n_{d,d,2}$ coincides with the closed Gromov-Witten invariants $n_{d}(h^{3})$ in Table \ref{oc_septicGW}.}
\label{openGW1}
\end{center}
\end{table}

As an example, we take $m=5$ and consider the septic Calabi-Yau fivefold. As demonstrated in \cite{Alim:2009bx} for the quintic Calabi-Yau threefold, in the vicinity of the large radius point $z_1=z_2 =0$, we can obtain two independent double logarithmic solutions as
\begin{align}
\begin{split}
F_1(t_1,t_2)&\equiv
\frac{6}{2}\frac{1}{(2\pi i)^2\Pi_0(z_1,z_2)}\frac{\partial^2}{\partial \rho_1^2}\left.\Pi(z_1,z_2;\rho_1,0)\right|_{\rho_1=0}
\\
&=
\frac{6}{2}t_1^2+\frac{1}{(2\pi i)^2}\mathop{\sum_{d_1,d_2=0}^{\infty}}\limits_{(d_1,d_2)\ne (0,0)}n_{d_1,d_2,1}{\rm Li}_2(q_1^{d_1}q_2^{d_2}),
\end{split}
\end{align}
and
\begin{align}
\begin{split}
F_2(t_1,t_2)&\equiv
\frac{1}{(2\pi i)^2\Pi_0(z_1,z_2)}
\Big(\frac{1}{2}\frac{\partial^2}{\partial \rho_1^2}
+\frac{1}{2}\frac{\partial^2}{\partial \rho_1\rho_2}
+\frac{7}{2}\frac{\partial^2}{\partial \rho_2^2}\Big)
\left.\Pi(z_1,z_2;\rho_1,\rho_2)\right|_{\rho_1=\rho_2=0}
\\
&=
\frac{1}{2}t_1^2+7t_1t_2+\frac{7}{2}t_2^2
+\frac{1}{(2\pi i)^2}\mathop{\sum_{d_1,d_2=0}^{\infty}}\limits_{(d_1,d_2)\ne (0,0)}n_{d_1,d_2,2}{\rm Li}_2(q_1^{d_1}q_2^{d_2}),
\end{split}
\end{align}
where we have used the mirror maps (\ref{mirror_m_Xm}) and $q_i=e^{2\pi i t_i}$. The coefficients $n_{d_1,d_2,1}$, $n_{d_1,d_2,2}$ are regarded as off-shell open Gromov-Witten invariants and indeed take integer values as listed in Table \ref{openGW1}. Note that in terms of Section \ref{sec:h_open_closed}, $F_1(t_1,t_2)$ and $F_1(t_1,t_2)+F_2(t_1,t_2)$ have informations of the two-point functions for $p=1$ in (\ref{two_point_X_comp}) represented by
\begin{equation}
\big<{\cal O}_{h_1}{\cal O}_{h_1^{4}}\big>_{\mathrm{\ICP}^1}^{\widehat{X}^c_{7}}\Big|_{{\rm Im}\; s \to \infty}\ \ \textrm{and} \ \
\big<{\cal O}_{h}{\cal O}_{h^{3}S}\big>_{\mathrm{\ICP}^1}^{\widehat{X}^c_{7}}\Big|_{{\rm Im}\; s \to \infty},
\end{equation}
respectively. Then from Table \ref{openGW1} we can confirm the relation for $p=1$ in (\ref{closed_two_X_comp}):
\begin{equation}
\big<{\cal O}_{h}{\cal O}_{h^{3}S}\big>_{\mathrm{\ICP}^1}^{\widehat{X}^c_{7}}\Big|_{{\rm Im}\; s \to \infty}=\big<{\cal O}_{h}{\cal O}_{h^{3}}\big>_{\mathrm{\ICP}^1}^{X_{7}}.
\end{equation}
In a similar way, we can also check for the case of $p=2$.


\end{document}